# Hybrid integration methods for on-chip quantum photonics

JE-HYUNG KIM[1], SHAHRIAR AGHAEIMEIBODI[2], JACQUES CAROLAN[3], DIRK ENGLUND[3*], AND EDO WAKS[2,4*]

[1]Department of Physics, Ulsan National Institute of Science and Technology (UNIST), Ulsan 44919, Republic of Korea
[2]Department of Electrical and Computer Engineering and Institute for Research in Electronics and Applied Physics, University of Maryland, College Park, Maryland 20742, United States
[3]Department of Electrical Engineering and Computer Science, Massachusetts Institute of Technology, Cambridge, Massachusetts 02139, United States
[4]Joint Quantum Institute, University of Maryland and the National Institute of Standards and Technology, College Park, Maryland 20742, United States
*Corresponding author: englund@mit.edu, edowaks@umd.edu

**Abstract.** The goal of integrated quantum photonics is to combine components for the generation, manipulation, and detection of non-classical light in a phase stable and efficient platform. Solid-state quantum emitters have recently reached outstanding performance as single photon sources. In parallel, photonic integrated circuits have been advanced to the point that thousands of components can be controlled on a chip with high efficiency and phase stability. Consequently, researchers are now beginning to combine these leading quantum emitters and photonic integrated circuit platforms to realize the best properties of each technology. In this article, we review recent advances in integrated quantum photonics based on such hybrid systems. Although hybrid integration solves many limitations of individual platforms, it also introduces new challenges that arise from interfacing different materials. We review various issues in solid-state quantum emitters and photonic integrated circuits, the hybrid integration techniques that bridge these two systems, and methods for chip-based manipulation of photons and emitters. Finally, we discuss the remaining challenges and future prospects of on-chip quantum photonics with integrated quantum emitters.

## 1. INTRODUCTION

The laws of quantum mechanics promise information processing technologies that are inherently more powerful than their classical counterparts, with examples including quantum computing [1], unconditionally secure communications [2], and quantum-enhanced precision sensing [3]. After decades of intensive theoretical and experimental efforts, the field of quantum information processing is reaching a critical stage: quantum computers and special-purpose quantum information processors may solve problems that classical computers cannot [4-6], and quantum networks can distribute entanglement over continental distances [7].

Photons are a promising system to realize quantum information processing applications due to their low noise properties, excellent modal control, and long-distance propagation [8]. These properties enable all-optical quantum technologies [9] and photonic interfaces between matter qubits [10]. By leveraging advances in photonic integrated circuits (PICs) for classical optical communications, integrated quantum photonics enables the chip-scale manipulation of quantum states of light, demonstrating orders of magnitude improvements in component density, loss, and phase stability compared to bulk-optical approaches. Such advances have enabled proof-of-principle demonstrations of quantum protocols, such as foundational tests of quantum mechanics [11], quantum simulation [12,13], and quantum machine learning [14]. Generally, such demonstrations are comprised of three distinct components — the generation of quantum states of light, their propagation through linear and nonlinear optical circuitry, and single photon readout. Bringing these components together into a single integrated system could enable a new generation of quantum optical processors capable of solving practical problems in quantum chemistry [15,16] and inference [17,18].

However, fully integrating the generation, manipulation, and detection of photons is an outstanding challenge for the field due to the unique material requirements for each distinct component. For example, epitaxially grown III-V semiconductor quantum dots are a leading approach for the near-deterministic generation of single photons in terms of purity, brightness, and indistinguishability [19]. However, the loss per component of III-V platforms is relatively high, and likely not at the level required for a large-scale photonic quantum technology [20]. In contrast, silicon photonics is unrivaled in terms of component density, scale, and compatibility with complementary metal–oxide–semiconductor (CMOS) electronics [21], with classical systems featuring over 1000 active components [22] and integration with millions of transistors [23]. Moreover, silicon-photonic-based quantum systems have demonstrated control of > 100 components [24] as well as the generation of entangled states of light [25]. However, methods to generate photons in silicon are based on spontaneous processes, such as four-wave mixing [25], or are incompatible with deterministic solid-state quantum emitters at visible or infrared wavelengths.

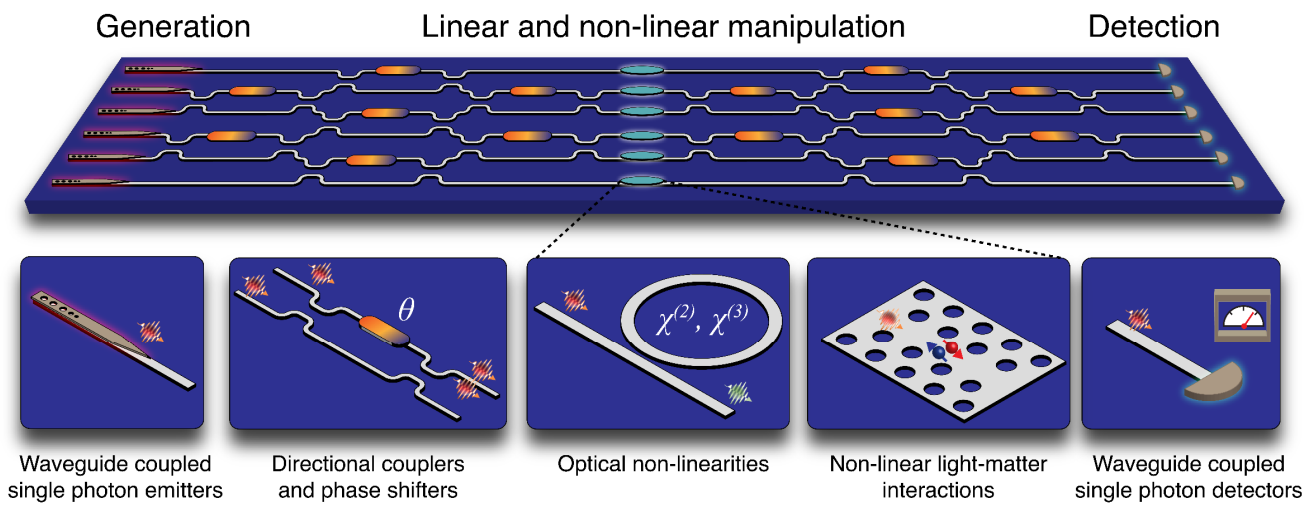

Fig. 1. Schematic of a hybrid integrated quantum photonic circuit consisting of different modules for the generation, linear and non-linear manipulation, and detection of non-classical light on a single chip. These individual modules are shown in more detail in the lower row of panels. Quantum emitters generate photons and route them to low-loss photonic waveguides. The combination of directional couplers and phase shifters enable arbitrary linear operations on the photons. The use of optical nonlinearities by resonant photonics (e.g., ring-resonators) as well as light-matter interactions expand the functionality of quantum photonics to the non-linear regime. Lastly, efficient on-chip single photon detectors can read-out the photons without the need for lossy photon extraction from the chip.

Hybrid integration provides a potential solution by incorporating disparate photonic technologies into a single integrated system that may not be otherwise compatible in a single fabrication process. Hybrid integration techniques include pick-and-place, wafer bonding, and epitaxial growth. In the context of quantum technologies, hybrid integration offers the tantalizing goal of bringing together quantum emitters, quantum memories, coherent linear and nonlinear operations, and single photon detection into a single quantum photonic platform as described in Fig. 1. In this paper, we review the emerging field of hybrid integration for next-generation quantum photonic processors, including platforms for quantum emitters and PICs, as well as techniques for their hybrid integration. Additionally, we explore on-chip methods for achieving coherent control of quantum photonic systems.

## 2. SOLID-STATE QUANTUM EMITTERS

Solid-state quantum emitters provide an essential building block for photon-based quantum technologies with their ability to produce single photons or entangled photon pairs in a deterministic manner [26,27]. To date, various types of solid-state quantum emitters, including quantum dots and atomic defects in crystals, have demonstrated single photon emissions with high purity and indistinguishability [19,28], as well as the potential for room-temperature operation [29-31] and compatibility with electrically-driven devices [32]. Also, their emission wavelength ranges from ultraviolet to near infra-red, which includes telecom wavelengths [31,33]. New solid-state quantum emitters are continually being reported in 2D materials [29,34] and perovskite nanocrystals [35], as well as for various crystal defects [30,36-38] [Fig. 2(a)].

Since the solid-state medium hosts the quantum emitters, they do not require a complicated trapping setup, which is essential for cold atoms and trapped ions. However, the solid-state environments create several issues such as limited light extraction efficiency, randomness of the position and frequency, and dephasing induced by interaction with charges and phonons in the quantum emitters. Initial efforts to solve these issues have focused on efficient generation of single photons by employing various micro/nanophotonic structures, including photonic crystals, photonic nanowires, microdisks, and micropillars [Fig. 2(b)]. Such structures dramatically improve the brightness of solid-state quantum emitters with enhanced light extraction [39,40], and also improve the collection efficiency and generation rate of single photons by far-field engineering [33,41], and Purcell enhancement [42,43]. Furthermore, researchers are continually developing techniques for controlling the emitters' position [44,45], frequency [46-48], and dephasing [28] [40], which have brought solid-state quantum emitters to the forefront of quantum light sources. Comprehensive reviews on solid-state emitters and important developments can be found in Ref. [26,49]

Recently, to achieve scalable and integrated quantum photonic systems, significant efforts have been made to realize monolithically or heterogeneously integrated quantum emitters with photonic circuits [Fig. 2(c)]. These on-chip integrated emitters serve as internal and deterministic quantum light sources for PICs. However, manipulating multiple quantum emitters in the photonic circuits poses new challenges. We discuss recent key developments and issues in on-chip integrated quantum emitters in photonic circuits in sections 4 and 5.

## 3. PHOTONIC INTEGRATED CIRCUIETS FOR QUANTUM PHOTONICS

PICs provide a compact, phase stable, and high-bandwidth platform to transmit, manipulate, and detect light on-chip. By leveraging advances in semiconductor manufacturing for classical communication, PICs have been demonstrated with over a thousand active components in a few square mm [50]. Now, with many foundries offering multi-project wafer runs in a variety of material platforms, the end-user can access complex PICs in a cost-effective manner, expanding the application areas of integrated photonics. Due to these favorable properties, PICs have emerged as a promising platform with which to generate and control quantum states of light at a scale required for practical optical quantum technologies [9,21]. In the context of hybrid integration, a PIC serves as a "photonic backbone" both to route and process single photons with high-fidelity, and to directly engineer the quantum emitter characteristics. When designing a photonic backbone, a number of key features should be considered, including loss budget, material compatibility, wavelength compatibility, manufacturability, modulation requirements, and power budget. In the following, we examine a number of such features.

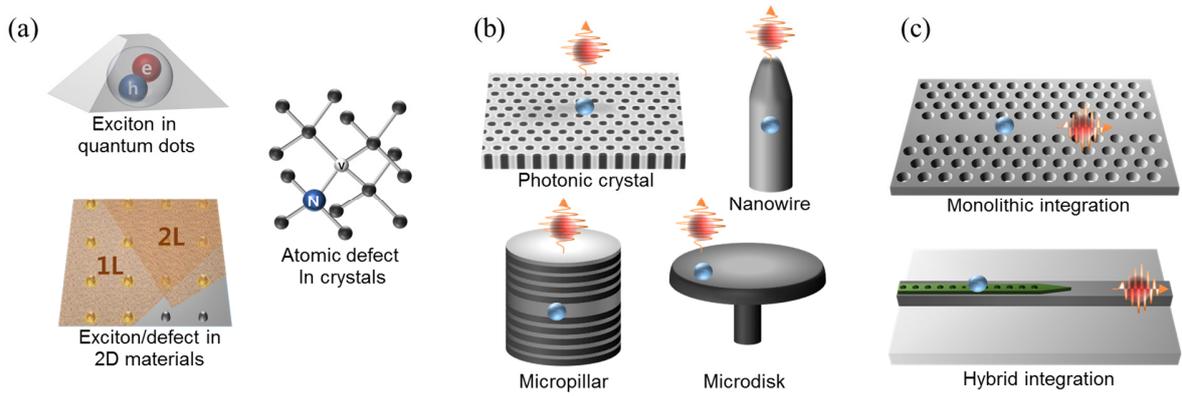

Fig. 2. (a) Various solid-state quantum emitters from single excitons in quantum dots, atomic defects in crystals, and exciton or defects in 2D materials. (b) Quantum emitters integrated with micro/nanophotonic structures, such as photonic crystals, photonic nanowires, micropillars, and microdisks. (c) Monolithic integration of the quantum emitter with an on-chip waveguide and the hybrid integration of the quantum emitter in a nanobeam (green color) on a heterogeneous photonic circuit. The blue and red spheres in (b) and (c) represent quantum emitters and single photons, respectively.

### A. Material Platforms

Many material platforms exist, each with varying levels of maturity. For example, silicon photonics benefits from an advanced silicon-on-insulator (SOI) manufacturing process that enables co-integration of photonics and CMOS electronics, enabling thousands of opto-electronic components on a single chip [24] [Fig. 3(a)]. Moreover, the high refractive index contrast between the Si core and $SiO_2$ cladding $\Delta n = (n_{\text{core}}^2 - n_{\text{clad}}^2)/2n_{\text{core}}^2 \approx 0.8$ enables compact componentry, which, alongside low propagation losses (as low as 2.7 dB/m [51]), enables low loss per component [21].

In the context of hybrid integration, one limitation of the SOI platform is a bandgap at ~1.1 μm, as many solid-state quantum emitters generate photons below this wavelength, causing significant loss. An approach for overcoming this limitation is to use telecom-compatible quantum emitters, such as InAs/InP quantum dots [33,52,53], defect centers in SiC [30] and GaN [54], and rare-earth-based quantum memories [55]. Moreover, the integration of these emitters into the SOI platform has been demonstrated [56]. Alternatively, one can move to a waveguide material with a higher bandgap energy. For example, $Si_3N_4$ is transparent above 400 nm, and low-pressure chemical vapor deposition techniques onto a $SiO_2$ layer provides a high-quality $Si_3N_4$ layer with precisely controlled thickness. The moderate index contrast $\Delta n = 0.25$, alongside low surface roughness, enables waveguides with ultra-low-losses of 0.1 dB/m [57] (at the cost of a larger bend radius and therefore greater device footprint), which is important for on-chip delay lines [58] [Fig. 3(b)]. Recently, $Si_3N_4$ has been included into the SOI foundry process enabling 3D integration [59].

In terms of emerging quantum photonic platforms, $LiNbO_3$ possesses strong electro-optic and acousto-optic properties [60,61] [Fig. 3(c)] and has a large transparency window of 350-4500 nm, making it appealing for hybrid integration. Due to the challenges in etching the material, initial efforts to develop waveguides in $LiNbO_3$ relied on titanium diffusion or proton exchange. However, the low refractive index contrast limited the scale of the devices [60]. More recently, advances in processing have enabled high-confinement nanophotonic waveguides fabricated from thin film $LiNbO_3$ on insulator, with losses as low as 2.7 dB/m [62] at telecom wavelengths and 6 dB/m at visible wavelengths [63]. Additionally, such waveguides have been integrated with quantum emitters [64]. AlN has also emerged as a promising platform for visible photonics [65], with a large transparency window [66] and modulation enabled by an intrinsic electro-optic [67] and piezoelectric effect [68]. Alternatively, III-V materials, such as InP, can enable the direct integration of active layers of quantum wells or quantum dots during the epitaxial growth process. III/V materials allow monolithic integration of light sources in photonic platforms [69][Fig. 3(d)]. However, compared to other materials, III-V-based PICs tend to have higher propagation loss around 2 dB/cm [70], and have a low bandgap energy which prohibits the use of visible light.

### B. Cryogenic Modulation

A key consideration for PICs for hybrid integration of quantum emitters is modulation at emitter-compatible cryogenic temperatures (< 10 K). Fast, low-loss modulation is critical for multiplexing [71], high-fidelity linear optical operations [72], and wavepacket engineering [73]. Materials with appreciable $\chi^{(2)}$ coefficients, such as $LiNbO_3$ [74] and AlN [65,74,75], enable switching via the Pockels effect, which is, in principle, not limited by cryogenic temperatures. Meanwhile, materials without an appreciable $\chi^{(2)}$ coefficient, such as Si or $Si_3N_4$ must rely on effects such as the plasma-dispersion effect [76], microelectromechanical effects [77], or thermo-optic effects [78]. Plasma-dispersion modulators, which rely on fast injection or depletion of carriers on fast timescales, have been demonstrated in Si microdisks at cryogenic temperatures [79], but the introduction of carriers causes loss which may be undesirable for quantum applications. Thermo-optic $Si_3N_4$ modulators have been demonstrated at cryogenic temperatures [80], however, the thermo-optic coefficient $dn/dT$ of both $Si_3N_4$ and $SiO_2$ decreases by an order of magnitude.

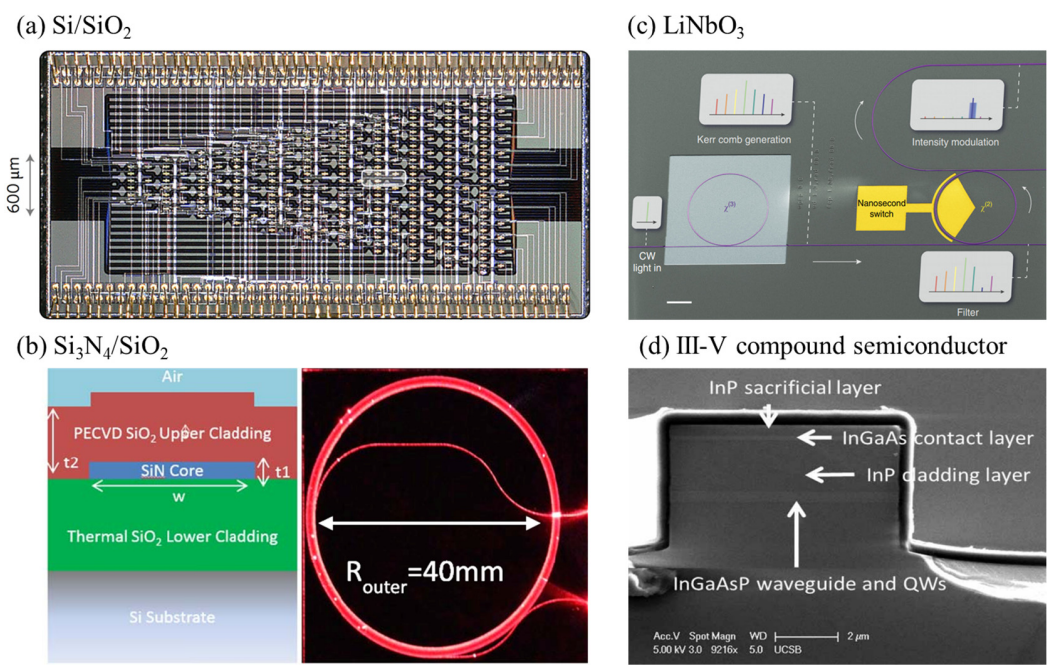

Fig. 3. (a) Optical image of a programmable Si PIC composed of 88 Mach-Zehnder interferometers, 26 input modes, 26 output modes, and 176 phase shifters [24]. (b) Top view of a $Si_3N_4$ waveguide coil (a 3 m-long spiral pattern) illuminated with a red laser [58]. (c) Scanning electron microscopy image of a $LiNbO_3$ photonic circuit consisting of a Kerr comb generator and an add-drop filter based on large $\chi^{(3)}$ and $\chi^{(2)}$ of $LiNbO_3$ [61]. (d) Cross-sectional scanning electron microscopy image of the epitaxial structure of an InGaAsP waveguide, including active quantum wells (QWs) [69].

An alternative approach is to incorporate materials with a strong Pockels effect into a non-electro-optically active material via hybrid integration. Organic polymers [81], $LiNbO_3$ [82], and electro-active oxides [83] have all been incorporated into Si. Notably, barium titanate possesses an exceptionally strong Pockels coefficient of 1000 pm/V at room temperature [84] and its integration with both Si and $Si_3N_4$ has been demonstrated at cryogenic temperatures, maintaining a Pockels coefficient of 200 pm/V [85] (compared with $LiNbO_3$ of 30 pm/V at room temperature).

Breakthroughs in hybrid integration of PICs for quantum photonics will benefit from a two-step approach: advances in PIC technology will open up new opportunities for hybrid integration, and fully understanding the unique requirements of quantum technologies will help direct PIC research.

## 4. HYBRID INTEGRATION TECHNOLOGY

PICs can efficiently manipulate and route light across the chip. To perform quantum information processing tasks, however, quantum light sources are required. These photons can be externally generated outside the chip and brought to it with various coupling techniques, or internally generated using the nonlinearity of the waveguide materials [86]. However, both approaches are currently falling short of the demanding efficiency requirements for complex quantum information processors [22]. A promising alternative is to integrate bright quantum emitters onto PICs directly. This could be beneficial for many aspects of the system, such as increased efficiency, scalability, stability, and controllability. However, creating a hybrid platform between the quantum emitter and the photonic circuit with efficient and deterministic coupling is a challenging task, and certain criteria must be considered. In this section, we review multiple techniques for hybrid integration and their ability to maintain high crystal quality and efficient optical coupling between the platforms, as well as their potential for scalability. The current state-of-the-art for hybrid integration of the quantum emitters onto photonic circuits is summarized in Table 1.

### A. Random dispersion

Quantum emitters in the form of nanoparticles, such as colloidal quantum dots or diamond nanoparticles, can be simply integrated with photonic structures by dispersing them onto the photonic platforms [87,88] [Fig. 4(a)]. Since the nanoparticle quantum emitters are not hosted in a dielectric medium, they can efficiently emit single photons without the problem of total internal reflection, a major issue for quantum emitters in a bulk medium. However, the nanoparticles themselves possess a large surface area, which often leads to optical instability, such as blinking or bleaching, due to the significant influence of the surface states and enhanced Auger process [89]. Therefore, additional surface treatment or environmental control may be required.

The simple dispersion method does not precisely control the position of the emitters, instead randomly places them near the photonic structures (e.g., waveguides or cavity structures). This fact limits the use of the random dispersion method for quantum photonic applications where the deterministic coupling of multiple quantum emitters with high coupling efficiency is crucial. To improve the coupling efficiency, it is possible to selectively disperse the nanoparticles using lithography-based masking [90] or tip manipulation of the particles in an atomic force microscope [91]. Therefore, with proper surface encapsulation and precise positioning techniques, this method could be an easy way to prototype and realize hybrid platforms.

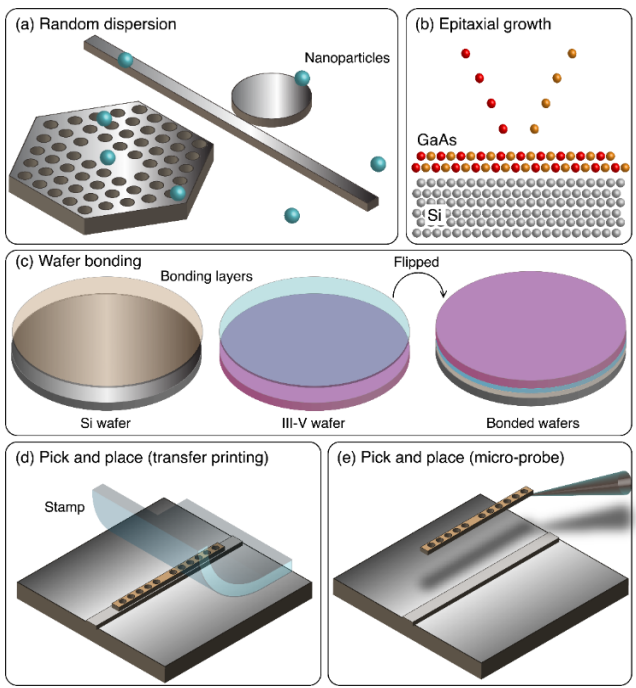

Fig. 4. Schematics of various hybrid integration methods for the quantum emitters on the photonic platforms. (a) Randomly dispersed nanoparticles in the vicinity of photonic structures, such as a microdisk or a photonic crystal cavity. (b) The epitaxial growth technique can be used to deposit layers such as GaAs on a Si substrate with a buffer layer (not shown). (c) Wafer-bonding technique to form a heterostructure of a III-V layer on a Si substrate. (d) Pick-and-place process by transfer printing a nanobeam containing quantum emitters on a waveguide, using a rubber stamp. (e) Pick-and-place process using a microprobe that places a nanobeam on a waveguide. Quantum emitters are embedded in the nano-structure.

**B. Epitaxial growth of hetero-structures**

Optically stable single photon emission with high single photon purity and indistinguishability can be generated from quantum emitters embedded in a high crystalline bulk medium, which can be achieved from epitaxially grown quantum dots or defects in a diamond film. Using the epitaxial growth technique, growing quantum materials directly on a photonic platform can provide hybrid hetero-structures for both emitters and photonic circuits in a single wafer. For example, hybrid hetero-structures of III-V compound semiconductors on Si, which are particularly important for realizing many optoelectronic applications [21,92], can be achieved using the epitaxial growth method [Fig. 4(b)]. However, growing such hetero-structures is not always favorable, often sacrificing the crystal quality due to the formation of antiphase boundaries and large mismatches in the materials' lattice constants, thermal coefficients, and charge polarity. To maintain crystal quality, a buffer layer needs to be inserted between the hetero-structures, and therefore, the quantum emitters require enough distance from the boundary, which reduces the coupling efficiency with the photonic circuits. Although the epitaxial growth of quantum materials on photonic circuits is still challenging, several new approaches, such as selective area growth and defect trapping, are being developed [93]. Therefore, this method still has strong potential for future on-chip hybrid quantum photonic devices.

**C. Wafer-bonding**

Another well-known method for integrating dissimilar material platforms is the wafer-to-wafer bonding technique, an example of which is shown in Fig. 4(c) [94]. Since each material is grown separately using its own optimized equipment and conditions, this method can maintain high crystal quality for both compounds and provide various material options that are more limited in the monolithic epitaxial growth technique. The wafer-bonding technique is also useful to couple the emission of quantum emitters to the photonic circuits since the emitters are placed on top surface of the transferred wafer with a thin capping layer, and the bonding process flips and bonds this top surface to the photonic wafer. Therefore, controlling the thickness of the capping layer of the emitters determines the distance between the emitter and the photonic circuits. The removal of the original substrate of the quantum emitters leaves a thin membrane structure on top of the PIC. With these hybrid hetero-structures, we can configure complicated electronic and photonic structures using micro/nanolithography techniques [62]. Figure 5(a) shows a quantum dot wafer orthogonally bonded to the side of a SiON photonic circuit, and Figure 5(b) shows a fabricated GaAs nanowire on a $Si_3N_4$ waveguide after the wafer bonding process of two wafers. One remaining problem for this technique is the random position and frequency of the emitters. Since the wafer-bonding method integrates two platforms on a wafer scale, without precise control of the position and frequency of the individual emitters, the actual coupling efficiency and yield remain low. However, recently developed techniques for site-controlled emitters [44,45], in-situ lithography [95,96], and local frequency tuning [46,62,97] may provide possible solutions for these issues. Figure 5(c) shows that the position of the quantum emitters in the bonded wafer is pre-defined by cathode luminescence in scanning electron microscopy, and the device is fabricated by in-situ electron beam lithography technique.

**D. Pick-and-place**

In the wafer-bonding technique, the independent growth of the materials for the quantum emitters and photonic platform preserves the crystal quality and provides hybrid hetero-structures at the wafer-scale. However, one limitation of this method is the reliance on random coupling between the emitters and photonic chips. To overcome this problem, a number of groups have suggested the pick-and-place method that transfers small-scale quantum devices one-by-one instead of wafer-scale integration. This single device transfer method allows the emitters to be pre-characterized before assembly [98,99], and therefore it is possible to selectively integrate desired emitters at a specific position of the photonic circuits. Another important feature of the pick-and-place method is that users are free to choose not only the materials but also the dimension and design of device structures for the emitters and photonic circuits, which is limited for pre-integrated wafers. Therefore, the two independently-designed systems can have more flexibility and functionality for controlling the emitters and photons on a chip. For example, the quantum emitters can be integrated with more complicated cavity structures to increase the radiative recombination rate and enhance the light-matter interaction strength [100]. Also, the pick-and-place technique can integrate various types of quantum materials, such as one-dimensional vertical nanowires [101,102] and two-dimensional van der Waals materials [103-105] that host the quantum emitters inside. This technique has also been successfully exploited to realize the integration of single photon detectors on a photonic circuit [106].

To detach the quantum emitter devices from the original wafer and release them onto pre-patterned photonic circuits, various methods have been demonstrated. A transfer printing method shown in Fig. 4(d) is one well-known example of the pick-and-place technique that uses an adhesive and transparent rubber stamp made of a material such as polydimethylsiloxane. With the pick-and-place method, assembling two pre-fabricated devices with high alignment accuracy is a crucial requirement since it significantly affects the coupling efficiency of the integrated emitters with the photonic chip. The use of transparent stamps enables the user to monitor the alignment in real-time with an optical microscope [see Fig. 5(d)], and additional alignment markers can increase the alignment accuracy [107,108]. In this case, the alignment accuracy is limited by the optical diffraction limit of around a few hundred nm for visible light. Another experimental challenge of this technique is the limited ability to re-position the emitters since the adhesion between the integrated structures is much stronger than their adhesion to the stamp. Therefore, the stamp cannot pick up the emitters again. Also, the stamping process tends to induce force over a large area and causes unwanted damage to the photonic circuit, such as physical damage on the photonic structures or detachment of the deposited metal electrodes. Introducing a carefully designed micro-stamp may avoid these problems [109] and could be used for highly integrated and fragile platforms.

Another effective technique for the pick-and-place approach is using a sharp micro-probe [56,101,102,110,111] [Fig. 4(e)]. A few micron or sub-micron-sized probe tip can pick up quantum emitters and transfer them onto the target position in either an optical [see Fig. 5(e)] or an electron microscope system [see Fig. 5(f)]. In particular, the latter environment significantly improves the alignment accuracy over the transfer printing method. Additionally, using the probe tip, it is possible to change the emitter position for better alignment accuracy even after integration. Furthermore, the sharp probe tip can pick up fragile single nanowires grown along the vertical direction [101,102] [Fig. 5(e)]. Even though handling quantum devices one-by-one with the pick-and-place technique requires a sophisticated process for the precise control of single devices, it provides the highest accuracy and controllability. Additionally, the process is compatible with various materials and structures. Further efforts for simplifying and automating the process may enable scalable and rapid fabrication of on-chip quantum photonic platforms with multiple deterministically integrated emitters.

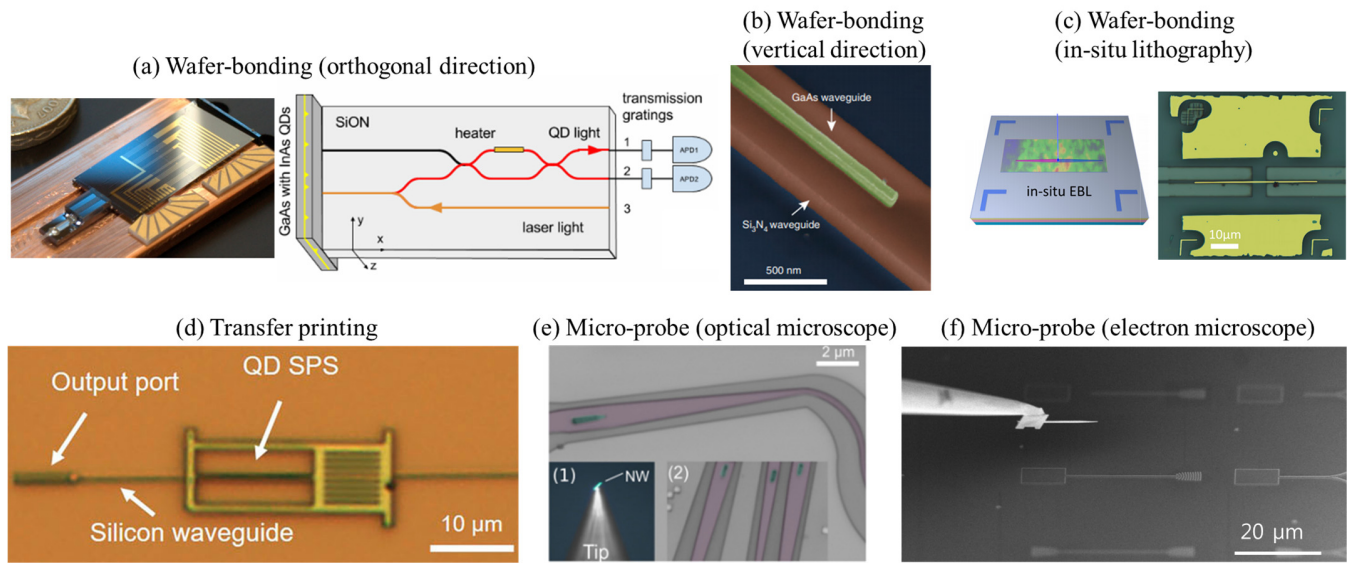

Fig. 5. Experimental demonstrations of hybrid integration of quantum emitters with photonic circuits using different integration techniques. (a) Optical image and schematics of integrated InAs quantum dots on a SiON photonic chip made by the orthogonal wafer-bonding method [112]. (b) A GaAs nanobeam on a $Si_3N_4$ waveguide by electron beam-lithography from a wafer-bonded GaAs/$Si_3N_4$ heterostructure [62]. (c) The left panel shows a schematic of in situ electron beam-lithography of a GaAs nanobeam aligned to a preselected QD. The right panel shows an optical microscopy image of fabricated devices (GaAs and $Si_3N_4$ colored in yellow and green, respectively) [96]. (d) Optical image of integrated InAs quantum dots (QDs) on a Si waveguide using a transfer printing method [107]. (e) Optical image of the transferred single nanowire-quantum dots on a $Si_3N_4$ waveguide using a microtip, with insets showing (1) picked nanowires (NW) on a tip and (2) integrated NWs on waveguides [102]. (f) Scanning electron microscopy image of an integrated InP nanobeam on a Si waveguide beamsplitter using a microprobe. Figure adapted from Ref. [56].

**Table 1. Comparative Summary of Representative Demonstrations with Integrated Quantum Emitters on a Photonic Chip**

| Integration Method | Quantum emitters | Photonic chip | Coupled emitters * | Coupling efficiency ** | $g^{(2)}(0)$ | Indistinguis hability | Detection | Demonstration *** |
|---|---|---|---|---|---|---|---|---|
| Wafer-bonding | Quantum dots | $Si_3N_4$ | 1 (gas-tuning) | 72 | 0.13 (with correction) | - | Fiber-coupled | Weak coupling [62] (microring resonator) |
| Wafer-bonding | Quantum dots | $Si_3N_4$ | 1 | 3 | 0.11 | 89 (At τ=0) | Fiber-coupled | Postselection using in-sutu lithography [113] |
| Wafer-bonding | Quantum dots | SiON | 20 (Stark tuning) | 8 | 0.23 | 54 (At τ=0) | Fiber-coupled | On-chip HOM [114] |
| Transfer printing | Quantum dots | GaAs | 2 | 63 | 0.23 | - | Free space (grating coupler) | Weak coupling [108] (nanobeam cavity) |
| Transfer printing | Quantum dots | SOI | 1 (temp. tuning) | 70 | 0.3 | - | Free space (grating coupler) | Weak coupling [107] (nanobeam cavity) |
| Transfer printing | $WSe_2$ | $LiNbO_3$ | 1 | 0.7 | - | - | Fiber-coupled | Waveguide coupling [105] |
| Micro probe | Quantum dots | SOI | 1 | 15 | 0.25 | - | Free space (grating coupler) | On-chip HBT [56] |
| Micro probe | Quantum dots | SOI | 1 (Stark tuning) | - | 0.12 | - | Free space (grating coupler) | Large frequency tuning [97] |
| Micro probe | Quantum dots | SOI | 1 (temp. tuning) | - | 0.25 | - | Free space (grating coupler) | On-chip frequency filtering [115] |
| Micro probe | Quantum dots | $LiNbO_3$ | 1 | - | 0.08 | - | Free space (grating coupler) | On-chip HBT [64] |
| Micro probe | Defect | $Si_3N_4$ | 1 | - | 0.07 (free space) 0.17 (on-chip) | - | Fiber-coupled | On-chip integration of quantum memory [99] |
| Micro probe | Quantum dots | $Si_3N_4$ | 1 (strain tuning) | 1 | 0.1 | - | Fiber-coupled | On-chip frequency tuning of emitters and ring-resonator [110] |

* The coupled emitters denote the number of studied or controlled emitters. The tuning mechanism is shown in parentheses.

** The coupling efficiency is determined between the quantum emitters and the waveguide.

***HBT and HOM represent Hanbury Brown and Twiss and Hong-Ou-Mandel interference experiments, respectively.

## 5. ON-CHIP CONTROL OF QUANTUM EMITTERS AND PHOTONS

Along with the efficient integration of quantum emitters with photonic circuits, controlling the quantum emitters to be identical to each other is essential to meet the criteria for quantum operation based on multiple, indistinguishable single photons. Furthermore, to establish efficient quantum operation on a chip, the photonic circuits should route, modulate, and detect the generated photons with minimal loss. In this section, we introduce the promising techniques for on-chip control of the emitters and photons as well as recent demonstrations of on-chip quantum operation.

### A. Coherent control of quantum emitters

Two-photon interference based on the Hong-Ou-Mandel interferometer is the primary mechanism for achieving measurement-based quantum interaction with photons [116]. The successful interference relies on highly coherent and indistinguishable single photons, which requires a sufficiently long coherence time ($\tau_2$) compared to the spontaneous decay time $\tau_1$, that is $\tau_2 \approx 2\tau_1$. However, the existence of phonon interactions and charge fluctuations in the solid-state environment causes timing and spectral jitters as well as pure dephasing, and thus the emitters have a broad emission linewidth compared to their intrinsic linewidth limited by the lifetime [117]. Such linewidth broadening is worse with an above-band excitation scheme that increases unnecessary interactions in solid-state systems. In the case of InAs quantum dots, the linewidth is typically over a few tens of $\mu$eV with the above-band excitation at a low temperature of 4 K, while their radiative decay time is as short as 1 ns, corresponding to a sub $\mu$eV homogeneous linewidth [118].

Recently, a number of groups have reported near transform-limited linewidth based on resonant [19,40] and quasi-resonant [119] methods. Figure 6(a) shows the indistinguishable visibility of quantum dots different excitation schemes: above-band [33,120], quasi-resonant [19,119,121,122], resonant [19,123-127], and two-photon excitations [128]. We note that increasing the degree of indistinguishability strongly depends on the excitation scheme, and high indistinguishability does not sacrifice the brightness in the quantum emitters, whereas heralded single photons from nonlinear processes have an inherent trade-off between the brightness and the indistinguishability [129]. Furthermore, driving the quantum emitters with a resonant laser shows interesting phenomena based on atom-photon interactions, such as Rabi oscillation and Mollow triplet [130], and also provides a way to control the quantum states of the emitters in a coherent manner, which is essential for quantum information processing [131].

One obstacle to the use of resonant excitation is a strong laser background scattered from the solid-state chip. Since the resonant scattered laser cannot be filtered out from the single photons using a spectral filter, it is necessary to employ other techniques for the separation of the two resonant signals. For example, a cross-polarization technique with a polarizing beam splitter combined with linear polarizers can selectively eliminate the laser background [132]. With an on-chip device, the nanophotonic waveguide can also act as a polarization filter [119,133-136]. Aligning the laser polarization direction along the waveguide direction prohibits the laser propagation in the waveguide [136]. Additionally, the large distance between the excitation and collection spots reduces the scattered laser signal further, as shown in Fig. 6(b). Employing a two-photon excitation method can also provide an alternative solution when the scattered laser light is unavoidable [28].

Together with phonon interaction, the fluctuating charge environment in the vicinity of the quantum emitters is another source of dephasing [137]. To stabilize the charge environment, surface passivation by adding a capping layer [138] or filling the charge trap with electrostatic field control [139] have been suggested.

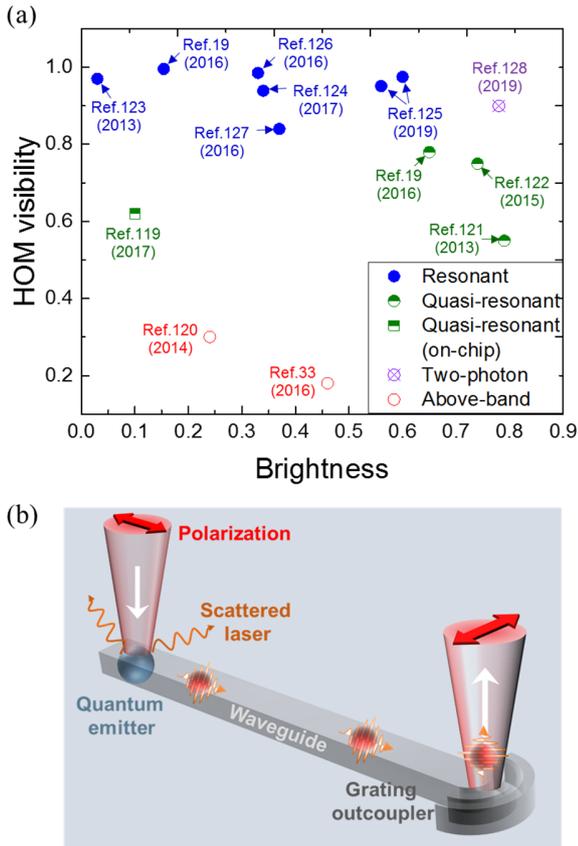

Fig. 6. (a) Comparison of the brightness and Hong-Ou-Mandel (HOM) interference visibility from quantum emitters driven with various excitation schemes: resonant (filled blue circles), quasi-resonant (half-filled green circles), quasi-resonant on-chip (green half-filled squares), two-photon (crossed purple circle), and above-band (empty red circles). Brightness is determined at the first lens or fiber. For the resonant excitation, the values consider the polarization optics, essential for suppressing the scattered laser and limiting the maximum brightness to 0.5 for unpolarized single photons. Quasi-resonant indicates that laser energy is lower than the wetting layer bandgap. (b) Schematic of resonant excitation of on-chip integrated quantum emitters in a nanophotonic waveguide that separates the single photons from the resonant excitation laser.

### B. Generation of multiple, indistinguishable single photons

Having coherent single photons from a single quantum emitter enables us to scale up to multiple indistinguishable single photon emitters on a chip. This is particularly important for large-scale photonic quantum simulators, such as boson samplers [140] and large-scale entangled photonic cluster states [141]. The most conventional way to produce multiple single photons is by parametric down-conversion in nonlinear media. However, this process is intrinsically probabilistic, and multiphoton events are inevitable as the brightness is increased. Therefore, the system becomes significantly inefficient with scale.

A bright single quantum emitter combined with a temporal-to-spatial demultiplexing technique is one possible way to achieve multiple single photons in a deterministic manner [Fig. 7(a)]. Multiple delay lines and beam splitters can spatially distribute the temporal array of single photons to multiple channels of the photonic circuit [124,142]. The advantage of this method is that the system only needs one bright single photon source with high purity and indistinguishability. For the deterministic distribution of the photons in each channel, electro-optic routing devices can be incorporated instead of passive beam splitters [142,143]. However, the degree of the indistinguishability decreases with the temporal separation between the photons [144]. Therefore, ultrafast electro-optic switches would be required to obtain maximum indistinguishability between photons. Furthermore, integrating a few tens of ns long delay lines for compensating the time interval between photons on a photonic circuit is a challenging task

Integrating multiple quantum emitters can offer a solution. The main challenge of incorporating multiple quantum emitters in a single chip is the frequency randomness of the quantum emitters, which limits quantum interference between photons from individual emitters. To eliminate this frequency mismatch between emitters, various local frequency tuning methods have been introduced. For example, Figure 7(b) shows quantum dots integrated into multiple channels of a SiON photonic circuit using wafer-bonding. The emission frequency of the integrated quantum emitters can be tuned independently by applied electric fields. Similar approaches have also been demonstrated in the InAs quantum dot-Si waveguide hybrid system [Fig. 7(c)] [97]. Another method of frequency tuning is by applying a local strain on the emitters. Within a hybrid system, this can be achieved by integrating the emitters on miniaturized piezoelectric actuator chips so that the platform can induce a local strain to individual emitters in an array [Fig. 7(d)] [46-48].

On-chip-integrated quantum emitters with matched frequency can provide not only multiple indistinguishable single photons, but also an outstanding platform to study many-body quantum physics. For example, multiple quantum emitters coupled to the same optical mode form entangled superposition states known as Dicke states, resulting in a super-radiant emission with an increased spontaneous emission rate [145]. In particular, integrating the emitters into a one-dimensional waveguide can realize long-range interactions between the emitters [146]. For example, Fig. 7(e) displays two far-separated quantum emitters coupled to a photonic crystal waveguide. When the frequencies of two emitters are tuned to resonance by local temperature tuning, they show superradiant emission as a result of collective behavior in a Dicke state. To date, various solid-state quantum systems have demonstrated such interaction on a chip [47,147,148], and recently, the super-radiance has been achieved with three quantum emitters in a waveguide with a local strain tuning method [47].

Along with the frequency control, the positional control of the emitters is another important factor for generating quantum emitter arrays with high coupling efficiency between the emitters and photonic circuits. Depending on the types of emitters, various experimental approaches have demonstrated deterministic positional control of the emitters. For instance, site-controlled quantum dots have been achieved by employing pre-patterned substrates or three-dimensional nanostructures such as pyramidal structures [44]. Growing vertical nanowires enables the placing of the single quantum dot in the middle of the nanowire, so the user can easily specify the position of the quantum dots during the growth process [149]. This nanowire structure is particularly useful for hybrid integration because it can control the position and number of quantum dots and be easily transferred into a photonic circuit with high coupling efficiency [102,110]. In the case of defects in crystals, ion-implantation techniques enable the control of the position and density of the defects by changing the dose value with the patterned mask [45]. Atomically thin 2D materials are also of great interest as arrayed single photon sources with their flexibility and tunability. For example, positioning quantum emitters with 2D materials can be achieved by transferring the material on nano-patterned substrates, which induce a local strain that can form strain-induced quantum emitters at deterministic positions [48,104,105].

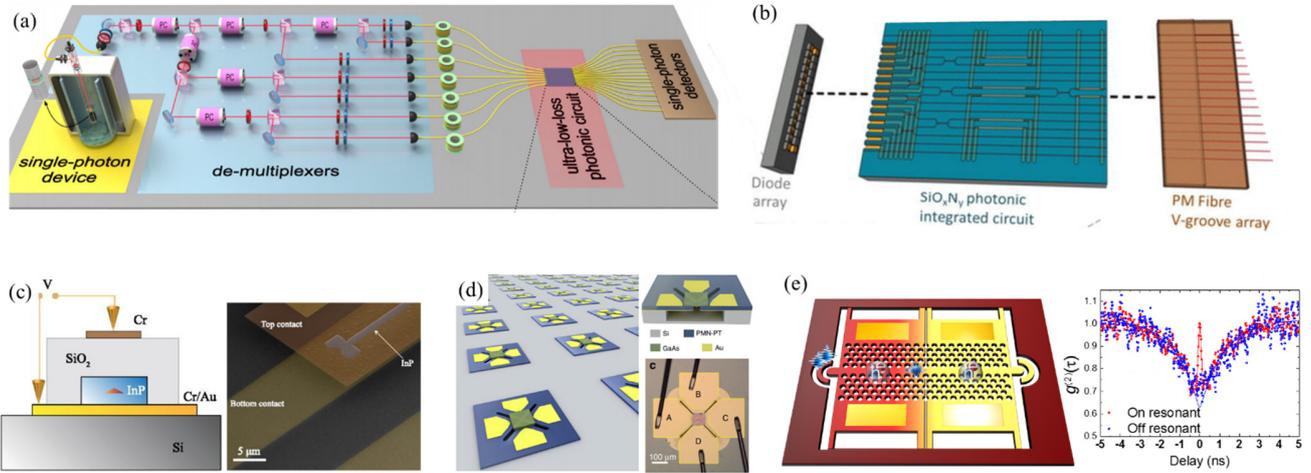

Fig. 7. (a) Schematic of the experimental setup for a boson sampler using a temporal-to-spatial demultiplexing technique with a single quantum dot [142]. (b) Independently tunable multiple quantum dot device integrated with a SiON photonic chip. Figure adapted from Ref. [114]. (c) Illustration and scanning electron microscopy image of the InAs quantum dot integrated with a Si substrate with a Stark tuning structure [97]. (d) Microelectromechanical systems for anisotropic strain engineering of quantum dot-based single photon sources [46]. (e) On-resonant two quantum dots in a photonic crystal waveguide with local heaters. The right panel shows superradiant emission as a result of the quantum interaction between two emitters coupled to a single optical mode of the waveguide [147].

These successful demonstrations of local controls of quantum emitters' frequency and position on a chip show strong potential of PICs with integrated sources of multiple identical quantum emitters and multiple indistinguishable single photons.

**C. On-chip manipulation of photons**

In the absence of direct photon-photon interaction, requiring strong Kerr nonlinearity, efficient quantum information processing can be established with quantum light sources, linear optical components, and detectors [150]. Using well-developed bulk or fiber optics such as mirrors, beam splitters, waveplates, and polarizers, we can easily manipulate the quantum state of photons to encode and decode the quantum information into the path, polarization, and time-bin of the photons. Realizing such optical components in PICs provides a promising solution for demonstrating a scalable and integrated quantum photonic system. Recent advances in PICs, as introduced in section 3, can highly integrate waveguides, beam splitters, phase shifters, and delay lines in a single chip. Combining these components can form tunable Mach-Zehnder interferometers, playing a key role in reconfigurable PICs [151].

The use of quantum emitters as quantum light sources requires additional photonic components to spectrally filter single photon emission from unwanted background emissions, including the scattered laser. Such frequency sorters have been demonstrated by using nano-photonic structures [101,115]. Figure 8(a) shows on-chip integrated quantum emitters and a micro-ring, acting as an add-drop filter. Such add-drop filters can sort out a narrow spectral line, and the resonant frequency can be tuned by controlling either the frequency of quantum emitters or the resonant mode of the add-drop filter.

The integrated nano-photonic structures add more functionality to PICs by enhancing optical nonlinearity. Frequency conversion of photons is one representative example and is very useful for the quantum emitters. Although, applying strain or an electric field on the quantum emitters can shift the emission frequency, the achievable tuning range typically remains below 10 nm. In contrast, the frequency conversion using $\chi^{(2)}$ or $\chi^{(3)}$ nonlinearity in the nano-photonic structures such as a ring resonator acts on photons and offers a much wider tuning range from a few tens of nm up to a few hundred nm. Figure 8(b) shows a waveguide-coupled resonator that converts the emission frequency of the quantum emitters using four-wave mixing Bragg scattering [152]. The fact that the frequency converter can match the emission frequency in a wide spectral range without local control of the emitters opens a new possibility of hybrid devices involving different types of quantum emitters, such as InAs quantum dots with near IR emission and defects in diamonds with visible emission in a chip. Such hybrid architecture will be very interesting because the system can provide efficient sources of photonic and spin qubits in the same chip, acting as quantum channels and memories, respectively.

## D. Spin-photon quantum interface

In the previous sections, we introduced on-chip generation and control of photons in PICs. Although photons provide an excellent carrier for quantum information, the storage time and deterministic interactions between photons are absent unless coupled to nonlinear matter. Integrated quantum emitters can provide not only photonic qubits but also spin qubits. Therefore, incorporating quantum-specific components, such as quantum memories and quantum gates, as well as coherent nonlinear optical elements based on stationary qubits, enable a wider range of photonic quantum information processing schemes [153] [Table 2] and new opportunities for exploiting quantum optics. For example, solid-state quantum emitters with a ground-state spin can mediate photon-photon interactions and store the information for a long time [154]. Recent advances in atomic defects in diamond have realized a coherent spin of over one second [155], and various new solid-state spins are emerging from several wide-bandgap semiconductors, such as SiC [156] and hBN [157].

Quantum entanglement between spins and photons has been demonstrated from various quantum emitters [131], and nanophotonic cavities or waveguides with strongly coupled quantum emitters can provide efficient spin-photon quantum interfaces by tailoring the light-matter interaction. In the context of cavity quantum electrodynamics (QED), the spin-photon interface controls the spin state via the polarization state of photons and vice versa. Recent work has demonstrated the conditional phase shift of photons [158], and strong photon-photon interaction [159] based on strongly coupled cavity-quantum emitter systems. Recent theoretical work also indicates that dynamically switchable cavities can mediate deterministic photon-photon gates with high fidelity [160]. Figure 8(c) shows an experimental demonstration of a single-photon switch using the spin of a charged quantum dot in a photonic crystal cavity. The result shows that nanophotonic structures with coupled solid-state spins can realize single-photon nonlinearity in a compact chip. Photon blockade is another example of strong nonlinearity at a single-photon level. The strongly coupled cavity-atomic system creates anharmonic ladder states that can alter the photon statistics from coherent to sub-Poissonian or super-Poissonian light sources and be used as photon number filters [161].

Along with the high $Q$ cavity, nanophotonic waveguides can also mediate efficient spin-photon interface using waveguide QED. The slow-light mode in the waveguide plays an important role in the waveguide QED, which has a similar principle to the cavity QED. Since the waveguides use propagation modes instead of localized modes, as in the cavity, multiple quantum emitters at different positions can couple to the waveguide and interact via real and virtual photons, enabling long-range connectivity. In addition, since the integrated emitter efficiently couples to the propagating photons in a waveguide, the emitter can induce strong optical nonlinearity at the single-photon level. Figure 8(d) displays an experimental demonstration of single-photon nonlinear optics in a waveguide that modifies the transmission of the waveguide with a coupled quantum emitter [162].

Therefore, the light-matter interaction with the integrated emitters enables a vast range of practical applications, such as quantum repeaters [163], quantum logic gates [158], photon-photon gates [164], single photon transistors [159], and photon number filters [165] in integrated photonic circuits. The studies show the potential capability of integrated quantum emitters as a source of photonic and stationary qubits and a mediator on spin-photon interfaces on a chip. However, so far, most demonstrations were performed in monolithically integrated platforms, which limits the number and function of the quantum emitters. The hybrid integration methods can provide a solution for scalable, integrated quantum photonic systems by the post-assembly of independently optimized emitters, cavities, and photonic circuits in a single chip. Recently, on-chip strong light-matter interaction in a hybrid system has been demonstrated with combined quantum emitters, high $Q$ cavity, and photonic waveguide as shown in Figure 8(e).

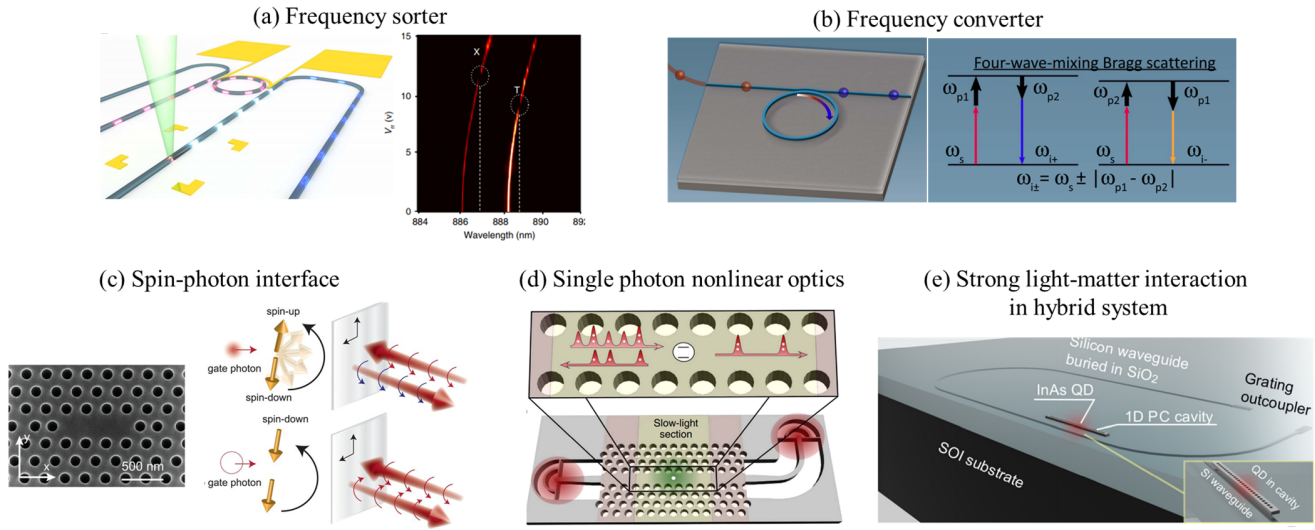

Fig. 8. (a) Frequency sorter based on a frequency tunable add-drop filter [101]. (b) Frequency converter using the four-wave mixing Bragg scattering process [152]. (c) Schematic image of a single-photon switch and transistor based on a single quantum dot in a photonic crystal cavity. The schematic shows that a gate photon controls the state of the spin, and then the spin determines the polarization of the signal field [159]. (d) Schematic image of controlled waveguide transmission with the coupled quantum emitter, showing a strong optical nonlinearity at a single-photon level [162]. (e) Demonstration of strong coupling between the quantum dot and the nanobeam high $Q$ cavity on a Si waveguide [166].

**Table 2. Representative Demonstrations of Quantum Took Kits for Integrated Quantum Photonic System**

| Quantum functional component | Role | Basic principle | Related work |
|---|---|---|---|
| Quantum memory | Store information in a photonic circuit | Long coherence time of spin | [99] |
| Spin-photon quantum interface | Control a spin (photon) state with a photon (spin) | Quantum entanglement between spins an photons | [131,158] |
| Photon-photon gate | Conditional photon switch | Strong optical nonlinear response mediated by emitters | [159] |
| Integrated quantum node | Large scale system involving multi-emitter coupling | Cooperative behavior of emitters mediated by photons | [47,147,148] |
| Photon number filter | Modification of photon statistics | Photon blockade using anharmonic ladder system | [161] |

### E. On-chip detection of photons

Quantum information processing ends with the efficient readout of the state of the photons. Since the photons travel along the waveguide in photonic circuits, to be detected it is necessary to extract out on-chip propagating photons and couple them into an objective lens or a fiber. To minimize the coupling loss, various methods have been suggested, such as grating-assisted coupling, evanescent coupling, tapered waveguides, and end-fiber coupling with a lensed fiber [167]. Although several schemes exist for efficient free space- and fiber-coupling, the coupling efficiency largely depends on the alignment and wavelength.

The most desirable way for detecting propagating photons in a chip is to integrate the detectors in the same chip. Single photon detectors based on superconducting nanowires are of great interest for this purpose because they can be fabricated on the photonic circuits directly and offer a fully integrated on-chip quantum photonic device, as shown in Fig. 9(a) [136,168]. Additionally, superconducting nanowire-based detectors outperform other detectors in terms of single photon detection characteristics, such as high efficiency of over 90%, fast response time below 50 ps, and high operation rates of over 100 MHz in a broad spectral range including the telecom wavelengths [169]. Furthermore, it is also possible to post-integrate separately fabricated detectors into the photonic circuits. For example, Figure 9(b) demonstrates the hybrid integration of the superconducting nanowire detector on a photonic waveguide using the pick-and-place technique [170].

## 6. REMAINING HURDLES AND OUTLOOK

In this review, we have presented recent advances in integrated quantum photonic systems that generate and manipulate quantum light and establish spin-photon interaction in a single chip. Solid-state quantum emitters now demonstrate high single photon generation rates, purities, and indistinguishability, with controlled position and frequency as well as spin with long coherence times. Meanwhile, photonic circuits can manipulate photons in various degrees of freedom using combined couplers, phase shifters, and linear/nonlinear components on a chip. Recent approaches for the hybrid integration of solid-state quantum emitters with photonic circuits have shown a possible solution for the long-standing issue of lack of internal, deterministic quantum light sources in the PICs. Also, integrating the quantum emitters in PICs provides many quantum functional components on a chip, and therefore it adds more functionality and flexibility for on-chip photonic quantum information processing.

However, despite this progress, realizing practical on-chip quantum photonic devices with integrated quantum emitters still faces many challenges. The principal obstacle is the need to generate multiple indistinguishable single photons from independently controlled quantum emitters. Although the number of quantum emitters that can be simultaneously controlled on a chip is increasing using several approaches, introduced in section 4, those emitters still lack long coherence times [47].

Another challenge is realizing efficient on-chip quantum interaction. We reviewed the possible mechanisms for such quantum interactions in section 5, which included two-photon interference using linear optics, atom-mediated nonlinear photon-photon interaction in cavities, and photon-mediated atom-atom interaction in waveguides. However, the interference visibility, single dipole cooperativity, and entanglement fidelity need to be further improved for a large-scale quantum system. To meet the performance criteria for deterministic quantum information processing with photons, higher efficiency, scalability, stability, and controllability of the emitter and photons are required. Satisfying all these conditions may be implausible within a single material. However, hybrid integration may pave the way by combining efficient single photon sources, coherent spins, and high-quality nanophotonic structures, as well as spectral and spatial control of the emitters and the photons.

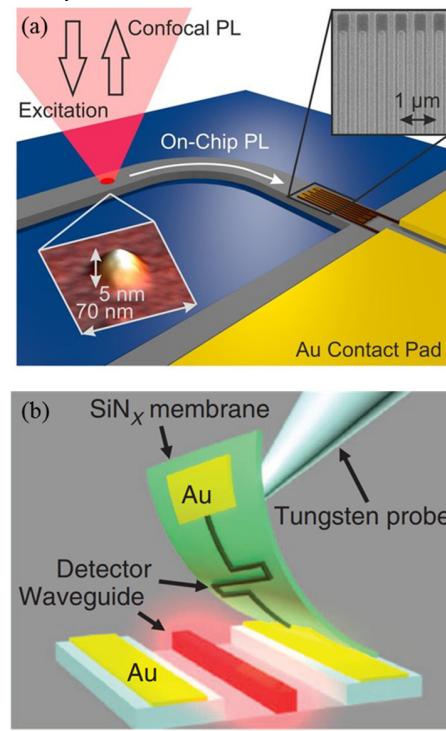

Fig. 9. (a) Schematic description of the on-chip detection of photons using an integrated superconducting nanowire detector [168]. (b) Integration of a single photon detector using a pick-and-place technique [170].

For applications, an electrically-driven system at room temperature is of great interest. Given the well-developed technology of semiconductor device manufacturing, electrically-driven single photon devices have been successfully demonstrated from various quantum emitters at room temperature [31,171,172]. Although those devices can efficiently generate single photons, the results have a lack of the indistinguishability of the single photons. To avoid significant spectral/timing jitters and dephasing induced by electrical excitation at the above bands, integrating a miniaturized tunable laser on the same chip has been suggested as a possible solution since it operates the system electrically but excites the quantum emitters optically at a resonant frequency [173,174]. For room-temperature operation, phonon interaction is unavoidable and broadens the emission linewidth, limiting indistinguishability. Therefore, achieving coherent single photons will be inherently difficult at high temperatures. However, recently, phonon decoupling in a low-dimensional system such as defects in two-dimensional hBN was reported, resulting in a Fourier transform-limited linewidth at room temperature [175].

Although it remains experimentally difficult to realize large-scale quantum photonic devices, the field of integrated quantum photonics is rising with developing quantum photonic technological capability, and it will provide a promising platform for various chip-scale quantum optics applications such as Boson sampling [176] and quantum chemistry [13] and also for large-scale photonic quantum processors enabling photonic cluster state quantum computing [141] and optical quantum networks [18,177]. Such integrated quantum photonic circuits can also interface with electronic microprocessors that can realize quantum-enhanced processing [23]. While quantum simulators and noisy intermediate-scale quantum processors are now becoming feasible [6,178], it is necessary to perform heuristic benchmarking on various problem classes. Large-scale systems with efficiently coupled spins and photons on a chip present a promising path to such applications.

**REFERENCES**

1. M. A. Nielsen, and I. Chuang, "Quantum Computation and Quantum Information," Am. J. Phys. **70**, 558-559 (2002)
2. N. Gisin, G. Ribordy, W. Tittel, and H. Zbinden, "Quantum cryptography," Reviews of Modern Physics **74**, 145-195 (2002)
3. V. Giovannetti, S. Lloyd, and L. Maccone, "Advances in quantum metrology," Nat. Photonics **5**, 222 (2011)
4. S. Aaronson, and A. Arkhipov, "Proceedings of the 43rd Annual ACM Symposium on Theory of Computing," 333 (2011)
5. A. Bouland, B. Fefferman, C. Nirkhe, and U. Vazirani, "On the complexity and verification of quantum random circuit sampling," Nat. Phys. **15**, 159-163 (2019)
6. F. Arute, K. Arya, R. Babbush, D. Bacon, J. C. Bardin, R. Barends, R. Biswas, S. Boixo, F. G. S. L. Brandao, D. A. Buell, B. Burkett, Y. Chen, Z. Chen, B. Chiaro, R. Collins, W. Courtney, A. Dunsworth, E. Farhi, B. Foxen, A. Fowler, C. Gidney, M. Giustina, R. Graff, K. Guerin, S. Habegger, M. P. Harrigan, M. J. Hartmann, A. Ho, M. Hoffmann, T. Huang, T. S. Humble, S. V. Isakov, E. Jeffrey, Z. Jiang, D. Kafri, K. Kechedzhi, J. Kelly, P. V. Klimov, S. Knysh, A. Korotkov, F. Kostritsa, D. Landhuis, M. Lindmark, E. Lucero, D. Lyakh, S. Mandrà, J. R. McClean, M. McEwen, A. Megrant, X. Mi, K. Michielsen, M. Mohseni, J. Mutus, O. Naaman, M. Neeley, C. Neill, M. Y. Niu, E. Ostby, A. Petukhov, J. C. Platt, C. Quintana, E. G. Rieffel, P. Roushan, N. C. Rubin, D. Sank, K. J. Satzinger, V. Smelyanskiy, K. J. Sung, M. D. Trevithick, A. Vainsencher, B. Villalonga, T. White, Z. J. Yao, P. Yeh, A. Zalcman, H. Neven, and J. M. Martinis, "Quantum supremacy using a programmable superconducting processor," Nature **574**, 505-510 (2019)
7. S.-K. Liao, W.-Q. Cai, J. Handsteiner, B. Liu, J. Yin, L. Zhang, D. Rauch, M. Fink, J.-G. Ren, W.-Y. Liu, Y. Li, Q. Shen, Y. Cao, F.-Z. Li, J.-F. Wang, Y.-M. Huang, L. Deng, T. Xi, L. Ma, T. Hu, L. Li, N.-L. Liu, F. Koidl, P. Wang, Y.-A. Chen, X.-B. Wang, M. Steindorfer, G. Kirchner, C.-Y. Lu, R. Shu, R. Ursin, T. Scheidl, C.-Z. Peng, J.-Y. Wang, A. Zeilinger, and J.-W. Pan, "Satellite-Relayed Intercontinental Quantum Network," Phys. Rev. Lett. **120**, 030501 (2018)
8. J. L. O'Brien, "Optical Quantum Computing," **318**, 1567-1570 (2007)
9. J. L. O'Brien, A. Furusawa, and J. Vučković, "Photonic quantum technologies," Nat. Photonics **3**, 687 (2009)
10. H. J. Kimble, "The quantum internet," Nature **453**, 1023-1030 (2008)
11. A. Peruzzo, P. Shadbolt, N. Brunner, S. Popescu, and J. L. O'Brien, "A Quantum Delayed-Choice Experiment," **338**, 634-637 (2012)
12. A. Aspuru-Guzik, and P. Walther, "Photonic quantum simulators," Nat. Phys. **8**, 285 (2012)
13. A. Peruzzo, J. McClean, P. Shadbolt, M.-H. Yung, X.-Q. Zhou, P. J. Love, A. Aspuru-Guzik, and J. L. O'Brien, "A variational eigenvalue solver on a photonic quantum processor," Nat. Commun. **5**, 4213 (2014)
14. J. Carolan, M. Mosheni, J. P. Olson, M. Prabhu, C. Chen, D. Bunandar, N. C. Harris, F. N. Wong, M. Hochberg, and S. J. a. p. a. Lloyd, "Variational Quantum Unsampling on a Quantum Photonic Processor," (2019)
15. J. Huh, G. G. Guerreschi, B. Peropadre, J. R. McClean, and A. Aspuru-Guzik, "Boson sampling for molecular vibronic spectra," Nat. Photonics **9**, 615 (2015)
16. C. Sparrow, E. Martín-López, N. Maraviglia, A. Neville, C. Harrold, J. Carolan, Y. N. Joglekar, T. Hashimoto, N. Matsuda, J. L. O'Brien, D. P. Tew, and A. Laing, "Simulating the vibrational quantum dynamics of molecules using photonics," Nature **557**, 660-667 (2018)
17. J. M. Arrazola, T. R. Bromley, J. Izaac, C. R. Myers, K. Brádler, and N. Killoran, "Machine learning method for state preparation and gate synthesis on photonic quantum computers," Quantum Sci. Technol. **4**, 024004 (2019)
18. G. R. Steinbrecher, J. P. Olson, D. Englund, and J. Carolan, "Quantum optical neural networks," npj Quantum Inf. **5**, 60 (2019)
19. N. Somaschi, V. Giesz, L. De Santis, J. C. Loredo, M. P. Almeida, G. Hornecker, S. L. Portalupi, T. Grange, C. Antón, J. Demory, C. Gómez, I. Sagnes, N. D. Lanzillotti-Kimura, A. Lemaître, A. Auffeves, A. G. White, L. Lanco, and P. Senellart, "Near-optimal single-photon sources in the solid state," Nat. Photonics **10**, 340 (2016)
20. T. Rudolph, "Why I am optimistic about the silicon-photonic route to quantum computing," APL Photonics **2**, 030901 (2017)
21. J. W. Silverstone, D. Bonneau, J. L. O'Brien, and M. G. Thompson, "Silicon Quantum Photonics," IEEE J. Sel. Top. Quantum Electron. **22**, 390-402 (2016)
22. S. Chung, H. Abediasl, and H. Hashemi, "15.4 A 1024-element scalable optical phased array in 0.18μm SOI CMOS," in *2017 IEEE International Solid-State Circuits Conference (ISSCC)*(2017), pp. 262-263.
23. C. Sun, M. T. Wade, Y. Lee, J. S. Orcutt, L. Alloatti, M. S. Georgas, A. S. Waterman, J. M. Shainline, R. R. Avizienis, S. Lin, B. R. Moss, R. Kumar, F. Pavanello, A. H. Atabaki, H. M. Cook, A. J. Ou, J. C. Leu, Y.-H. Chen, K. Asanović, R. J. Ram, M. A. Popović, and V. M. Stojanović, "Single-chip microprocessor that communicates directly using light," Nature **528**, 534 (2015)
24. N. C. Harris, G. R. Steinbrecher, M. Prabhu, Y. Lahini, J. Mower, D. Bunandar, C. Chen, F. N. C. Wong, T. Baehr-Jones, M. Hochberg, S. Lloyd, and D. Englund, "Quantum transport simulations in a programmable nanophotonic processor," Nat. Photonics **11**, 447 (2017)
25. J. W. Silverstone, D. Bonneau, K. Ohira, N. Suzuki, H. Yoshida, N. Iizuka, M. Ezaki, C. M. Natarajan, M. G. Tanner, R. H. Hadfield, V. Zwiller, G. D. Marshall, J. G. Rarity, J. L. O'Brien, and M. G. Thompson, "On-chip quantum interference between silicon photon-pair sources," Nat. Photonics **8**, 104 (2013)

26. I. Aharonovich, D. Englund, and M. Toth, "Solid-state single-photon emitters," Nat. Photonics **10**, 631 (2016)

27. H. Wang, H. Hu, T. H. Chung, J. Qin, X. Yang, J. P. Li, R. Z. Liu, H. S. Zhong, Y. M. He, X. Ding, Y. H. Deng, Q. Dai, Y. H. Huo, S. Höfling, C.-Y. Lu, and J.-W. Pan, "On-Demand Semiconductor Source of Entangled Photons Which Simultaneously Has High Fidelity, Efficiency, and Indistinguishability," Phys. Rev. Lett. **122**, 113602 (2019)

28. M. Reindl, K. D. Jöns, D. Huber, C. Schimpf, Y. Huo, V. Zwiller, A. Rastelli, and R. Trotta, "Phonon-Assisted Two-Photon Interference from Remote Quantum Emitters," Nano Lett. **17**, 4090-4095 (2017)

29. N. V. Proscia, Z. Shotan, H. Jayakumar, P. Reddy, C. Cohen, M. Dollar, A. Alkauskas, M. Doherty, C. A. Meriles, and V. M. Menon, "Near-deterministic activation of room-temperature quantum emitters in hexagonal boron nitride," Optica **5**, 1128-1134 (2018)

30. J. Wang, Y. Zhou, Z. Wang, A. Rasmita, J. Yang, X. Li, H. J. von Bardeleben, and W. Gao, "Bright room temperature single photon source at telecom range in cubic silicon carbide," Nat. Commun. **9**, 4106 (2018)

31. M. J. Holmes, K. Choi, S. Kako, M. Arita, and Y. Arakawa, "Room-Temperature Triggered Single Photon Emission from a III-Nitride Site-Controlled Nanowire Quantum Dot," Nano Lett. **14**, 982-986 (2014)

32. A. J. Bennett, P. Atkinson, P. See, M. B. Ward, R. M. Stevenson, Z. L. Yuan, D. C. Unitt, D. J. P. Ellis, K. Cooper, D. A. Ritchie, and A. J. Shields, "Single-photon-emitting diodes: a review," **243**, 3730-3740 (2006)

33. J.-H. Kim, T. Cai, C. J. K. Richardson, R. P. Leavitt, and E. Waks, "Two-photon interference from a bright single-photon source at telecom wavelengths," Optica **3**, 577-584 (2016)

34. S. Ren, Q. Tan, and J. Zhang, "Review on the quantum emitters in two-dimensional materials," J. Semicond. **40**, 071903 (2019)

35. H. Utzat, W. Sun, A. E. K. Kaplan, F. Krieg, M. Ginterseder, B. Spokoyny, N. D. Klein, K. E. Shulenberger, C. F. Perkinson, M. V. Kovalenko, and M. G. Bawendi, "Coherent single-photon emission from colloidal lead halide perovskite quantum dots," **363**, 1068-1072 (2019)

36. Y. Zhou, Z. Wang, A. Rasmita, S. Kim, A. Berhane, Z. Bodrog, G. Adamo, A. Gali, I. Aharonovich, and W.-b. Gao, "Room temperature solid-state quantum emitters in the telecom range," **4**, eaar3580 (2018)

37. H. Utzat, W. Sun, A. E. K. Kaplan, F. Krieg, M. Ginterseder, B. Spokoyny, N. D. Klein, K. E. Shulenberger, C. F. Perkinson, M. V. Kovalenko, and M. G. Bawendi, "Coherent single-photon emission from colloidal lead halide perovskite quantum dots," Science **363**, 1068-1072 (2019)

38. T. Iwasaki, F. Ishibashi, Y. Miyamoto, Y. Doi, S. Kobayashi, T. Miyazaki, K. Tahara, K. D. Jahnke, L. J. Rogers, B. Naydenov, F. Jelezko, S. Yamasaki, S. Nagamachi, T. Inubushi, N. Mizuochi, and M. Hatano, "Germanium-Vacancy Single Color Centers in Diamond," Sci. Rep. **5**, 12882 (2015)

39. M. E. Reimer, G. Bulgarini, N. Akopian, M. Hocevar, M. B. Bavinck, M. A. Verheijen, E. P. A. M. Bakkers, L. P. Kouwenhoven, and V. Zwiller, "Bright single-photon sources in bottom-up tailored nanowires," Nat. Commun. **3**, 737 (2012)

40. X. Ding, Y. He, Z. C. Duan, N. Gregersen, M. C. Chen, S. Unsleber, S. Maier, C. Schneider, M. Kamp, S. Höfling, C.-Y. Lu, and J.-W. Pan, "On-Demand Single Photons with High Extraction Efficiency and Near-Unity Indistinguishability from a Resonantly Driven Quantum Dot in a Micropillar," Phys. Rev. Lett. **116**, 020401 (2016)

41. N. Livneh, M. G. Harats, D. Istrati, H. S. Eisenberg, and R. Rapaport, "Highly Directional Room-Temperature Single Photon Device," Nano Lett. **16**, 2527-2532 (2016)

42. M. D. Birowosuto, H. Sumikura, S. Matsuo, H. Taniyama, P. J. van Veldhoven, R. Nötzel, and M. Notomi, "Fast Purcell-enhanced single photon source in 1,550-nm telecom band from a resonant quantum dot-cavity coupling," Sci. Rep. **2**, 321 (2012)

43. J. L. Zhang, S. Sun, M. J. Burek, C. Dory, Y.-K. Tzeng, K. A. Fischer, K. Kelaita, K. G. Lagoudakis, M. Radulaski, Z.-X. Shen, N. A. Melosh, S. Chu, M. Lončar, and J. Vučković, "Strongly Cavity-Enhanced Spontaneous Emission from Silicon-Vacancy Centers in Diamond," Nano Lett. **18**, 1360-1365 (2018)

44. G. Juska, V. Dimastrodonato, L. O. Mereni, A. Gocalinska, and E. Pelucchi, "Towards quantum-dot arrays of entangled photon emitters," Nat. Photonics **7**, 527 (2013)

45. Y.-I. Sohn, S. Meesala, B. Pingault, H. A. Atikian, J. Holzgrafe, M. Gündoğan, C. Stavrakas, M. J. Stanley, A. Sipahigil, J. Choi, M. Zhang, J. L. Pacheco, J. Abraham, E. Bielejec, M. D. Lukin, M. Atatüre, and M. Lončar, "Controlling the coherence of a diamond spin qubit through its strain environment," Nat. Commun. **9**, 2012 (2018)

46. Y. Chen, J. Zhang, M. Zopf, K. Jung, Y. Zhang, R. Keil, F. Ding, and O. G. Schmidt, "Wavelength-tunable entangled photons from silicon-integrated III–V quantum dots," Nat. Commun. **7**, 10387 (2016)

47. J. Q. Grim, A. S. Bracker, M. Zalalutdinov, S. G. Carter, A. C. Kozen, M. Kim, C. S. Kim, J. T. Mlack, M. Yakes, B. Lee, and D. Gammon, "Scalable in operando strain tuning in nanophotonic waveguides enabling three-quantum-dot superradiance," Nat. Mater. **18**, 963-969 (2019)

48. H. Kim, J. S. Moon, G. Noh, J. Lee, and J.-H. Kim, "Position and Frequency Control of Strain-Induced Quantum Emitters in WSe2 Monolayers," Nano Lett. **19**, 7534-7539 (2019)

49. P. Senellart, G. Solomon, and A. White, "High-performance semiconductor quantum-dot single-photon sources," Nat. Nanotechnol. **12**, 1026 (2017)

50. S. Chung, H. Abediasl, and H. Hashemi, "A Monolithically Integrated Large-Scale Optical Phased Array in Silicon-on-Insulator CMOS," IEEE Journal of Solid-State Circuits **53**, 275-296 (2018)

51. A. Biberman, M. J. Shaw, E. Timurdogan, J. B. Wright, and M. R. Watts, "Ultralow-loss silicon ring resonators," in *The 9th International Conference on Group IV Photonics (GFP)*(2012), pp. 39-41.

52. X. Liu, K. Akahane, N. A. Jahan, N. Kobayashi, M. Sasaki, H. Kumano, and I. Suemune, "Single-photon emission in telecommunication band from an InAs quantum dot grown on InP with molecular-beam epitaxy," Appl. Phys. Lett. **103**, 061114 (2013)

53. M. Benyoucef, M. Yacob, J. P. Reithmaier, J. Kettler, and P. Michler, "Telecom-wavelength (1.5 μm) single-photon emission from InP-based quantum dots," Appl. Phys. Lett. **103**, 162101 (2013)

54. Y. Zhou, Z. Wang, A. Rasmita, S. Kim, A. Berhane, Z. Bodrog, G. Adamo, A. Gali, I. Aharonovich, and W.-b. Gao, "Room temperature solid-state quantum emitters in the telecom range," Science Advances **4**, eaar3580 (2018)

55. A. M. Dibos, M. Raha, C. M. Phenicie, and J. D. Thompson, "Atomic Source of Single Photons in the Telecom Band," Phys. Rev. Lett. **120**, 243601 (2018)

56. J.-H. Kim, S. Aghaeimeibodi, C. J. K. Richardson, R. P. Leavitt, D. Englund, and E. Waks, "Hybrid Integration of Solid-State Quantum Emitters on a Silicon Photonic Chip," Nano Lett. **17**, 7394-7400 (2017)

57. J. F. Bauters, M. J. R. Heck, D. John, D. Dai, M.-C. Tien, J. S. Barton, A. Leinse, R. G. Heideman, D. J. Blumenthal, and J. E. Bowers, "Ultra-low-loss high-aspect-ratio Si3N4 waveguides," Opt. Express **19**, 3163-3174 (2011)

58. T. Huffman, M. Davenport, M. Belt, J. E. Bowers, and D. J. Blumenthal, "Ultra-Low Loss Large Area Waveguide Coils for Integrated Optical Gyroscopes," IEEE Photonics Technology Letters **29**, 185-188 (2017)

59. W. D. Sacher, J. C. Mikkelsen, P. Dumais, J. Jiang, D. Goodwill, X. Luo, Y. Huang, Y. Yang, A. Bois, P. G.-Q. Lo, E. Bernier, and J. K. S. Poon, "Tri-layer silicon nitride-on-silicon photonic platform for ultra-low-loss crossings and interlayer transitions," Opt. Express **25**, 30862-30875 (2017)

60. M. Bazzan, and C. Sada, "Optical waveguides in lithium niobate: Recent developments and applications," Appl. Phys. Rev. **2**, 040603 (2015)

61. C. Wang, M. Zhang, M. Yu, R. Zhu, H. Hu, and M. Loncar, "Monolithic lithium niobate photonic circuits for Kerr frequency comb generation and modulation," Nat. Commun. **10**, 978 (2019)

62. M. Davanco, J. Liu, L. Sapienza, C.-Z. Zhang, J. V. De Miranda Cardoso, V. Verma, R. Mirin, S. W. Nam, L. Liu, and K. Srinivasan, "Heterogeneous integration for on-chip quantum photonic circuits with single quantum dot devices," Nat. Commun. **8**, 889 (2017)

63. B. Desiatov, A. Shams-Ansari, M. Zhang, C. Wang, and M. Lončar, "Ultra-low-loss integrated visible photonics using thin-film lithium niobate," Optica **6**, 380-384 (2019)

64. S. Aghaeimeibodi, B. Desiatov, J.-H. Kim, C.-M. Lee, M. A. Buyukkaya, A. Karasahin, C. J. K. Richardson, R. P. Leavitt, M. Lončar, and E. Waks, "Integration of quantum dots with lithium niobate photonics," Appl. Phys. Lett. **113**, 221102 (2018)

65. T.-J. Lu, M. Fanto, H. Choi, P. Thomas, J. Steidle, S. Mouradian, W. Kong, D. Zhu, H. Moon, K. Berggren, J. Kim, M. Soltani, S. Preble, and D. Englund, "Aluminum nitride integrated photonics platform for the ultraviolet to visible spectrum," Opt. Express **26**, 11147-11160 (2018)

66. N. Watanabe, T. Kimoto, and J. Suda, "The temperature dependence of the refractive indices of GaN and AlN from room temperature up to 515 °C," J. Appl. Phys. **104**, 106101 (2008)

67. C. Xiong, W. H. P. Pernice, and H. X. Tang, "Low-Loss, Silicon Integrated, Aluminum Nitride Photonic Circuits and Their Use for Electro-Optic Signal Processing," Nano Lett. **12**, 3562-3568 (2012)

68. S. A. Tadesse, and M. Li, "Sub-optical wavelength acoustic wave modulation of integrated photonic resonators at microwave frequencies," Nat. Commun. **5**, 5402 (2014)

69. M. Lu, H. Park, E. Bloch, A. Sivananthan, A. Bhardwaj, Z. Griffith, L. A. Johansson, M. J. Rodwell, and L. A. Coldren, "Highly integrated optical heterodyne phase-locked loop with phase/frequency detection," Opt. Express **20**, 9736-9741 (2012)

70. M. Smit, X. Leijtens, H. Ambrosius, E. Bente, J. van der Tol, B. Smalbrugge, T. de Vries, E.-J. Geluk, J. Bolk, R. van Veldhoven, L. Augustin, P. Thijs, D. D'Agostino, H. Rabbani, K. Lawniczuk, S. Stopinski, S. Tahvili, A. Corradi, E. Kleijn, D. Dzibrou, M. Felicetti, E. Bitincka, V. Moskalenko, J. Zhao, R. Santos, G. Gilardi, W. Yao, K. Williams, P. Stabile, P. Kuindersma, J. Pello, S. Bhat, Y. Jiao, D. Heiss, G. Roelkens, M. Wale, P. Firth, F. Soares, N. Grote, M. Schell, H. Debregeas, M. Achouche, J.-L. Gentner, A. Bakker, T. Korthorst, D. Gallagher, A. Dabbs, A. Melloni, F. Morichetti, D. Melati, A. Wonfor, R. Penty, R. Broeke, B. Musk, and D. Robbins, "An introduction to InP-based generic integration technology," Semicond. Sci. Technol. **29**, 083001 (2014)

71. F. Kaneda, and P. G. Kwiat, "High-efficiency single-photon generation via large-scale active time multiplexing," Science Advances **5**, eaaw8586 (2019)

72. J. Carolan, C. Harrold, C. Sparrow, E. Martín-López, N. J. Russell, J. W. Silverstone, P. J. Shadbolt, N. Matsuda, M. Oguma, M. Itoh, G. D. Marshall, M. G. Thompson, J. C. F. Matthews, T. Hashimoto, J. L. O'Brien, and A. Laing, "Universal linear optics," Science **349**, 711-716 (2015)

73. M. Karpiński, M. Jachura, L. J. Wright, and B. J. Smith, "Bandwidth manipulation of quantum light by an electro-optic time lens," Nat. Photonics **11**, 53 (2016)

74. C. Wang, M. Zhang, X. Chen, M. Bertrand, A. Shams-Ansari, S. Chandrasekhar, P. Winzer, and M. Lončar, "Integrated lithium niobate electro-optic modulators operating at CMOS-compatible voltages," Nature **562**, 101-104 (2018)

75. S. Zhu, and G.-Q. Lo, "Aluminum nitride electro-optic phase shifter for backend integration on silicon," Opt. Express **24**, 12501-12506 (2016)

76. D. J. Thomson, F. Y. Gardes, J. Fedeli, S. Zlatanovic, Y. Hu, B. P. P. Kuo, E. Myslivets, N. Alic, S. Radic, G. Z. Mashanovich, and G. T. Reed, "50-Gb/s Silicon Optical Modulator," IEEE Photonics Technology Letters **24**, 234-236 (2012)

77. T. J. Seok, N. Quack, S. Han, R. S. Muller, and M. C. Wu, "Large-scale broadband digital silicon photonic switches with vertical adiabatic couplers," Optica **3**, 64-70 (2016)

78. N. C. Harris, Y. Ma, J. Mower, T. Baehr-Jones, D. Englund, M. Hochberg, and C. Galland, "Efficient, compact and low loss thermo-optic phase shifter in silicon," Opt. Express **22**, 10487-10493 (2014)

79. M. Gehl, C. Long, D. Trotter, A. Starbuck, A. Pomerene, J. B. Wright, S. Melgaard, J. Siirola, A. L. Lentine, and C. DeRose, "Operation of high-speed silicon photonic micro-disk modulators at cryogenic temperatures," Optica **4**, 374-382 (2017)

80. A. W. Elshaari, I. E. Zadeh, K. D. Jöns, and V. Zwiller, "Thermo-Optic Characterization of Silicon Nitride Resonators for Cryogenic Photonic Circuits," IEEE Photonics Journal **8**, 1-9 (2016)

81. M. Lauermann, S. Wolf, P. C. Schindler, R. Palmer, S. Koeber, D. Korn, L. Alloatti, T. Wahlbrink, J. Bolten, M. Waldow, M. Koenigsmann, M. Kohler, D. Malsam, D. L. Elder, P. V. Johnston, N. Phillips-Sylvain, P. A. Sullivan, L. R. Dalton, J. Leuthold, W. Freude, and C. Koos, "40 GBd 16QAM Signaling at 160 Gb/s in a Silicon-Organic Hybrid Modulator," J. Lightwave Technol. **33**, 1210-1216 (2015)

82. M. He, M. Xu, Y. Ren, J. Jian, Z. Ruan, Y. Xu, S. Gao, S. Sun, X. Wen, L. Zhou, L. Liu, C. Guo, H. Chen, S. Yu, L. Liu, and X. Cai, "High-performance hybrid silicon and lithium niobate Mach–Zehnder modulators for 100 Gbit s−1 and beyond," Nat. Photonics **13**, 359-364 (2019)

83. S. Abel, F. Eltes, J. E. Ortmann, A. Messner, P. Castera, T. Wagner, D. Urbonas, A. Rosa, A. M. Gutierrez, D. Tulli, P. Ma, B. Baeuerle, A. Josten, W. Heni, D. Caimi, L. Czornomaz, A. A. Demkov, J. Leuthold, P. Sanchis, and J. Fompeyrine, "Large Pockels effect in micro- and nanostructured barium titanate integrated on silicon," Nat. Mater. **18**, 42-47 (2019)

84. S. Abel, D. Caimi, M. Sousa, T. Stöferle, C. Rossel, C. Marchiori, A. Chelnokov, and J. Fompeyrine, *Electro-optical properties of barium titanate films epitaxially grown on silicon* (SPIE, 2012).

85. F. Eltes, G. E. Villarreal-Garcia, D. Caimi, H. Siegwart, A. A. Gentile, A. Hart, P. Stark, G. D. Marshall, M. G. Thompson, and J. Barreto, "An integrated cryogenic optical modulator," arXiv preprint arXiv:1904.10902 (2019)

86. L. Caspani, C. Xiong, B. J. Eggleton, D. Bajoni, M. Liscidini, M. Galli, R. Morandotti, and D. J. Moss, "Integrated sources of photon quantum states based on nonlinear optics," Light: Science & Applications **6**, e17100-e17100 (2017)

87. Z. Yang, M. Pelton, I. Fedin, D. V. Talapin, and E. Waks, "A room temperature continuous-wave nanolaser using colloidal quantum wells," Nat. Commun. **8**, 143 (2017)

88. C. Santori, P. E. Barclay, K. M. C. Fu, R. G. Beausoleil, S. Spillane, and M. Fisch, "Nanophotonics for quantum optics using nitrogen-vacancy centers in diamond," Nanotechnology **21**, 274008 (2010)

89. A. L. Efros, and D. J. Nesbitt, "Origin and control of blinking in quantum dots," Nat. Nanotechnol. **11**, 661 (2016)

90. Y. Chen, A. Ryou, M. R. Friedfeld, T. Fryett, J. Whitehead, B. M. Cossairt, and A. Majumdar, "Deterministic Positioning of Colloidal Quantum Dots on Silicon Nitride Nanobeam Cavities," Nano Lett. **18**, 6404-6410 (2018)

91. F. Böhm, N. Nikolay, C. Pyrlik, J. Schlegel, A. Thies, A. Wicht, G. Tränkle, and O. Benson, "On-chip integration of single solid-state quantum emitters with a SiO2 photonic platform," New J. Phys. **21**, 045007 (2019)

92. N. C. Harris, D. Bunandar, M. Pant, G. R. Steinbrecher, J. Mower, M. Prabhu, T. Baehr-Jones, M. Hochberg, and D. Englund, "Large-scale quantum photonic circuits in silicon," Nanophotonics **5**, 456-468 (2016)

93. B. Kunert, Y. Mols, M. Baryshniskova, N. Waldron, A. Schulze, and R. Langer, "How to control defect formation in monolithic III/V hetero-epitaxy on (100) Si: A critical review on current approaches," Semicond. Sci. Technol. **33**, 093002 (2018)

94. K. Tanabe, K. Watanabe, and Y. Arakawa, "III-V/Si hybrid photonic devices by direct fusion bonding," Sci. Rep. **2**, 349 (2012)

95. P. Schnauber, J. Schall, S. Bounouar, T. Höhne, S.-I. Park, G.-H. Ryu, T. Heindel, S. Burger, J.-D. Song, S. Rodt, and S. Reitzenstein, "Deterministic Integration of Quantum Dots into on-Chip Multimode Interference Beamsplitters Using in Situ Electron Beam Lithography," Nano Lett. **18**, 2336-2342 (2018)

96. P. Schnauber, A. Singh, J. Schall, S. I. Park, J. D. Song, S. Rodt, K. Srinivasan, S. Reitzenstein, and M. Davanco, "Indistinguishable Photons from Deterministically Integrated Single Quantum Dots in Heterogeneous GaAs/Si3N4 Quantum Photonic Circuits," Nano Lett. **19**, 7164-7172 (2019)

97. S. Aghaeimeibodi, C.-M. Lee, M. A. Buyukkaya, C. J. K. Richardson, and E. Waks, "Large stark tuning of InAs/InP quantum dots," Appl. Phys. Lett. **114**, 071105 (2019)

98. K. Mnaymneh, D. Dalacu, J. McKee, J. Lapointe, S. Haffouz, J. F. Weber, D. B. Northeast, P. J. Poole, G. C. Aers, and R. L. Williams, "On-Chip Integration of Single Photon Sources via Evanescent Coupling of Tapered Nanowires to SiN Waveguides," Adv. Quantum Technol. **0**, 1900021

99. S. L. Mouradian, T. Schröder, C. B. Poitras, L. Li, J. Goldstein, E. H. Chen, M. Walsh, J. Cardenas, M. L. Markham, D. J. Twitchen, M. Lipson, and D. Englund, "Scalable Integration of Long-Lived Quantum Memories into a Photonic Circuit," Phys. Rev. X **5**, 031009 (2015)

100. R. Katsumi, Y. Ota, A. Osada, T. Tajiri, T. Yamaguchi, M. Kakuda, S. Iwamoto, H. Akiyama, and Y. Arakawa, "In-situ wavelength tuning of quantum-dot single-photon sources integrated on a CMOS silicon chip," arXiv preprint arXiv:1909.13452 (2019)

101. A. W. Elshaari, I. E. Zadeh, A. Fognini, M. E. Reimer, D. Dalacu, P. J. Poole, V. Zwiller, and K. D. Jöns, "On-chip single photon filtering and multiplexing in hybrid quantum photonic circuits," Nat. Commun. **8**, 379 (2017)

102. I. E. Zadeh, A. W. Elshaari, K. D. Jöns, A. Fognini, D. Dalacu, P. J. Poole, M. E. Reimer, and V. Zwiller, "Deterministic Integration of Single Photon Sources in Silicon Based Photonic Circuits," Nano Lett. **16**, 2289-2294 (2016)

103. P. Tonndorf, O. Del Pozo-Zamudio, N. Gruhler, J. Kern, R. Schmidt, A. I. Dmitriev, A. P. Bakhtinov, A. I. Tartakovskii, W. Pernice, S. Michaelis de Vasconcellos, and R. Bratschitsch, "On-Chip Waveguide Coupling of a Layered Semiconductor Single-Photon Source," Nano Lett. **17**, 5446-5451 (2017)

104. F. Peyskens, C. Chakraborty, M. Muneeb, D. Van Thourhout, and D. Englund, "Integration of Single Photon Emitters in 2D Layered Materials with a Silicon Nitride Photonic Chip," arXiv preprint arXiv:1904.08841 (2019)

105. D. White, A. Branny, R. J. Chapman, R. Picard, M. Brotons-Gisbert, A. Boes, A. Peruzzo, C. Bonato, and B. D. Gerardot, "Atomically-thin quantum dots integrated with lithium niobate photonic chips [Invited]," Opt. Mater. Express **9**, 441-448 (2019)

106. F. Najafi, J. Mower, N. C. Harris, F. Bellei, A. Dane, C. Lee, X. Hu, P. Kharel, F. Marsili, S. Assefa, K. K. Berggren, and D. Englund, "On-chip detection of non-classical light by scalable integration of single-photon detectors," Nat. Commun. **6**, 5873 (2015)

107. R. Katsumi, Y. Ota, A. Osada, T. Yamaguchi, T. Tajiri, M. Kakuda, S. Iwamoto, H. Akiyama, and Y. Arakawa, "Quantum-dot single-photon source on a CMOS silicon photonic chip integrated using transfer printing," APL Photonics **4**, 036105 (2019)

108. R. Katsumi, Y. Ota, M. Kakuda, S. Iwamoto, and Y. Arakawa, "Transfer-printed single-photon sources coupled to wire waveguides," Optica **5**, 691-694 (2018)

109. J. Lee, I. Karnadi, J. T. Kim, Y.-H. Lee, and M.-K. Kim, "Printed Nanolaser on Silicon," ACS Photonics **4**, 2117-2123 (2017)

110. A. W. Elshaari, E. Büyüközer, I. E. Zadeh, T. Lettner, P. Zhao, E. Schöll, S. Gyger, M. E. Reimer, D. Dalacu, P. J. Poole, K. D. Jöns, and V. Zwiller, "Strain-Tunable Quantum Integrated Photonics," Nano Lett. **18**, 7969-7976 (2018)

111. L. Li, I. Bayn, M. Lu, C.-Y. Nam, T. Schröder, A. Stein, N. C. Harris, and D. Englund, "Nanofabrication on unconventional substrates using transferred hard masks," Sci. Rep. **5**, 7802 (2015)

112. E. Murray, D. J. P. Ellis, T. Meany, F. F. Floether, J. P. Lee, J. P. Griffiths, G. A. C. Jones, I. Farrer, D. A. Ritchie, A. J. Bennett, and A. J. Shields, "Quantum photonics hybrid integration platform," Appl. Phys. Lett. **107**, 171108 (2015)

113. P. Schnauber, A. Singh, J. Schall, S. I. Park, J. D. Song, S. Rodt, K. Srinivasan, S. Reitzenstein, and M. Davanco, "Indistinguishable Photons from Deterministically Integrated Single Quantum Dots in Heterogeneous GaAs/Si3N4 Quantum Photonic Circuits," Nano Lett. (2019)

114. D. J. P. Ellis, A. J. Bennett, C. Dangel, J. P. Lee, J. P. Griffiths, T. A. Mitchell, T.-K. Paraiso, P. Spencer, D. A. Ritchie, and A. J. Shields, "Independent indistinguishable quantum light sources on a reconfigurable photonic integrated circuit," Appl. Phys. Lett. **112**, 211104 (2018)

115. S. Aghaeimeibodi, J.-H. Kim, C.-M. Lee, M. A. Buyukkaya, C. Richardson, and E. Waks, "Silicon photonic add-drop filter for quantum emitters," Opt. Express **27**, 16882-16889 (2019)

116. W. B. Gao, P. Fallahi, E. Togan, A. Delteil, Y. S. Chin, J. Miguel-Sanchez, and A. Imamoğlu, "Quantum teleportation from a propagating photon to a solid-state spin qubit," Nat. Commun. **4**, 2744 (2013)

117. P. Tighineanu, C. L. Dreeßen, C. Flindt, P. Lodahl, and A. S. Sørensen, "Phonon Decoherence of Quantum Dots in Photonic Structures: Broadening of the Zero-Phonon Line and the Role of Dimensionality," Phys. Rev. Lett. **120**, 257401 (2018)

118. J. Liu, K. Konthasinghe, M. Davanço, J. Lawall, V. Anant, V. Verma, R. Mirin, S. W. Nam, J. D. Song, B. Ma, Z. S. Chen, H. Q. Ni, Z. C. Niu, and K. Srinivasan, "Single Self-Assembled InAs/GaAs Quantum Dots in Photonic Nanostructures: The Role of Nanofabrication," Phys. Rev. Appl. **9**, 064019 (2018)

119. G. Kiršanskė, H. Thyrrestrup, R. S. Daveau, C. L. Dreeßen, T. Pregnolato, L. Midolo, P. Tighineanu, A. Javadi, S. Stobbe, R. Schott, A. Ludwig, A. D. Wieck, S. I. Park, J. D. Song, A. V. Kuhlmann, I. Söllner, M. C. Löbl, R. J. Warburton, and P. Lodahl, "Indistinguishable and efficient single photons from a quantum dot in a planar nanobeam waveguide," Phys. Rev. B **96**, 165306 (2017)

120. X. Liu, H. Kumano, H. Nakajima, S. Odashima, T. Asano, T. Kuroda, and I. Suemune, "Two-photon interference and coherent control of single InAs quantum dot emissions in an Ag-embedded structure," J. Appl. Phys. **116**, 043103 (2014)

121. O. Gazzano, S. Michaelis de Vasconcellos, C. Arnold, A. Nowak, E. Galopin, I. Sagnes, L. Lanco, A. Lemaître, and P. Senellart, "Bright solid-state sources of indistinguishable single photons," Nat. Commun. **4**, 1425 (2013)

122. V. Giesz, S. Portalupi, T. Grange, C. Antón, L. De Santis, J. Demory, N. Somaschi, I. Sagnes, A. Lemaître, and L. Lanco, "Cavity-enhanced two-photon interference using remote quantum dot sources," Phys. Rev. B **92**, 161302 (2015)

123. Y.-M. He, Y. He, Y.-J. Wei, D. Wu, M. Atatüre, C. Schneider, S. Höfling, M. Kamp, C.-Y. Lu, and J.-W. Pan, "On-demand semiconductor single-photon source with near-unity indistinguishability," Nat. Nanotechnol. **8**, 213 (2013)

124. H. Wang, Y. He, Y.-H. Li, Z.-E. Su, B. Li, H.-L. Huang, X. Ding, M.-C. Chen, C. Liu, J. Qin, J.-P. Li, Y.-M. He, C. Schneider, M. Kamp, C.-Z. Peng, S. Höfling, C.-Y. Lu, and J.-W. Pan, "High-efficiency multiphoton boson sampling," Nat. Photonics **11**, 361 (2017)

125. H. Wang, Y.-M. He, T. H. Chung, H. Hu, Y. Yu, S. Chen, X. Ding, M. C. Chen, J. Qin, X. Yang, R.-Z. Liu, Z. C. Duan, J. P. Li, S. Gerhardt, K. Winkler, J. Jurkat, L.-J. Wang, N. Gregersen, Y.-H. Huo, Q. Dai, S. Yu, S. Höfling, C.-Y. Lu, and J.-W. Pan, "Towards optimal single-photon sources from polarized microcavities," Nat. Photonics **13**, 770-775 (2019)

126. X. Ding, Y. He, Z.-C. Duan, N. Gregersen, M.-C. Chen, S. Unsleber, S. Maier, C. Schneider, M. Kamp, and S. Höfling, "On-demand single photons with high extraction efficiency and near-unity indistinguishability from a resonantly driven quantum dot in a micropillar," Phys. Rev. Lett. **116**, 020401 (2016)

127. S. Unsleber, Y.-M. He, S. Gerhardt, S. Maier, C.-Y. Lu, J.-W. Pan, N. Gregersen, M. Kamp, C. Schneider, and S. Höfling, "Highly indistinguishable on-demand resonance fluorescence photons from a deterministic quantum dot micropillar device with 74% extraction efficiency," Opt. Express **24**, 8539-8546 (2016)

128. H. Wang, H. Hu, T.-H. Chung, J. Qin, X. Yang, J.-P. Li, R.-Z. Liu, H.-S. Zhong, Y.-M. He, and X. Ding, "On-demand semiconductor source of entangled photons which simultaneously has high fidelity, efficiency, and indistinguishability," Phys. Rev. Lett. **122**, 113602 (2019)

129. L. Yang, X. Ma, X. Guo, L. Cui, and X. Li, "Characterization of a fiber-based source of heralded single photons," Phys. Rev. A **83**, 053843 (2011)

130. E. B. Flagg, A. Muller, J. W. Robertson, S. Founta, D. G. Deppe, M. Xiao, W. Ma, G. J. Salamo, and C. K. Shih, "Resonantly driven coherent oscillations in a solid-state quantum emitter," Nat. Phys. **5**, 203-207 (2009)

131. W. B. Gao, A. Imamoglu, H. Bernien, and R. Hanson, "Coherent manipulation, measurement and entanglement of individual solid-state spins using optical fields," Nat. Photonics **9**, 363 (2015)

132. A. V. Kuhlmann, J. Houel, D. Brunner, A. Ludwig, D. Reuter, A. D. Wieck, and R. J. Warburton, "A dark-field microscope for background-free detection of resonance fluorescence from single semiconductor quantum dots operating in a set-and-forget mode," Rev. Sci. Instrum. **84**, 073905 (2013)

133. Ł. Dusanowski, S.-H. Kwon, C. Schneider, and S. Höfling, "Near-Unity Indistinguishability Single Photon Source for Large-Scale Integrated Quantum Optics," Phys. Rev. Lett. **122**, 173602 (2019)

134. M. N. Makhonin, J. E. Dixon, R. J. Coles, B. Royall, I. J. Luxmoore, E. Clarke, M. Hugues, M. S. Skolnick, and A. M. Fox, "Waveguide Coupled Resonance Fluorescence from On-Chip Quantum Emitter," Nano Lett. **14**, 6997-7002 (2014)

135. F. Liu, A. J. Brash, J. O'Hara, L. M. P. P. Martins, C. L. Phillips, R. J. Coles, B. Royall, E. Clarke, C. Bentham, N. Prtljaga, I. E. Itskevich, L. R. Wilson, M. S. Skolnick, and A. M. Fox, "High Purcell factor generation of indistinguishable on-chip single photons," Nat. Nanotechnol. **13**, 835-840 (2018)

136. M. Schwartz, E. Schmidt, U. Rengstl, F. Hornung, S. Hepp, S. L. Portalupi, K. Ilin, M. Jetter, M. Siegel, and P. Michler, "Fully On-Chip Single-Photon Hanbury-Brown and Twiss Experiment on a Monolithic Semiconductor–Superconductor Platform," Nano Lett. **18**, 6892-6897 (2018)

137. J. Houel, A. V. Kuhlmann, L. Greuter, F. Xue, M. Poggio, B. D. Gerardot, P. A. Dalgarno, A. Badolato, P. M. Petroff, A. Ludwig, D. Reuter, A. D. Wieck, and R. J. Warburton, "Probing Single-Charge Fluctuations at a $\mathrm{GaAs}/\mathrm{AlAs}$ Interface Using Laser Spectroscopy on a Nearby InGaAs Quantum Dot," Phys. Rev. Lett. **108**, 107401 (2012)

138. J. Liu, K. Konthasinghe, M. Davanço, J. Lawall, V. Anant, V. Verma, R. Mirin, S. W. Nam, J. D. Song, B. Ma, Z. S. Chen, H. Q. Ni, Z. C. Niu, and K. Srinivasan, "Single Self-Assembled $\mathrm{InAs}/\mathrm{GaAs}$ Quantum Dots in Photonic Nanostructures: The Role of Nanofabrication," Physical Review Applied **9**, 064019 (2018)

139. A. Reigue, A. Lemaître, C. G. Carbonell, C. Ulysse, K. Merghem, S. Guilet, R. Hostein, and V. Voliotis, "Resonance fluorescence revival in a voltage-controlled semiconductor quantum dot," Appl. Phys. Lett. **112**, 073103 (2018)

140. A. P. Lund, M. J. Bremner, and T. C. Ralph, "Quantum sampling problems, BosonSampling and quantum supremacy," npj Quantum Inf. **3**, 15 (2017)

141. M. Gimeno-Segovia, T. Rudolph, and S. E. Economou, "Deterministic Generation of Large-Scale Entangled Photonic Cluster State from Interacting Solid State Emitters," Phys. Rev. Lett. **123**, 070501 (2019)

142. H. Wang, W. Li, X. Jiang, Y. M. He, Y. H. Li, X. Ding, M. C. Chen, J. Qin, C. Z. Peng, C. Schneider, M. Kamp, W. J. Zhang, H. Li, L. X. You, Z. Wang, J. P. Dowling, S. Höfling, C.-Y. Lu, and J.-W. Pan, "Toward Scalable Boson Sampling with Photon Loss," Phys. Rev. Lett. **120**, 230502 (2018)

143. F. Lenzini, B. Haylock, J. C. Loredo, R. A. Abrahão, N. A. Zakaria, S. Kasture, I. Sagnes, A. Lemaitre, H.-P. Phan, D. V. Dao, P. Senellart, M. P. Almeida, A. G. White, and M. Lobino, "Active demultiplexing of single photons from a solid-state source," Laser Photonics Rev. **11**, 1600297 (2017)

144. A. Thoma, P. Schnauber, M. Gschrey, M. Seifried, J. Wolters, J. H. Schulze, A. Strittmatter, S. Rodt, A. Carmele, A. Knorr, T. Heindel, and S. Reitzenstein, "Exploring Dephasing of a Solid-State Quantum Emitter via Time- and Temperature-Dependent Hong-Ou-Mandel Experiments," Phys. Rev. Lett. **116**, 033601 (2016)

145. A. V. Andreev, V. I. Emel'yanov, and Y. A. Il'inskiĭ, "Collective spontaneous emission (Dicke superradiance)," Soviet Physics Uspekhi **23**, 493-514 (1980)

146. P. Solano, P. Barberis-Blostein, F. K. Fatemi, L. A. Orozco, and S. L. Rolston, "Super-radiance reveals infinite-range dipole interactions through a nanofiber," Nat. Commun. **8**, 1857 (2017)

147. J.-H. Kim, S. Aghaeimeibodi, C. J. K. Richardson, R. P. Leavitt, and E. Waks, "Super-Radiant Emission from Quantum Dots in a Nanophotonic Waveguide," Nano Lett. **18**, 4734-4740 (2018)

148. A. Sipahigil, R. E. Evans, D. D. Sukachev, M. J. Burek, J. Borregaard, M. K. Bhaskar, C. T. Nguyen, J. L. Pacheco, H. A. Atikian, C. Meuwly, R. M. Camacho, F. Jelezko, E. Bielejec, H. Park, M. Lončar, and M. D. Lukin, "An integrated diamond nanophotonics platform for quantum-optical networks," Science **354**, 847-850 (2016)

149. S. Deshpande, J. Heo, A. Das, and P. Bhattacharya, "Electrically driven polarized single-photon emission from an InGaN quantum dot in a GaN nanowire," Nat. Commun. **4**, 1675 (2013)

150. E. Knill, R. Laflamme, and G. J. Milburn, "A scheme for efficient quantum computation with linear optics," Nature **409**, 46-52 (2001)

151. C. Harris Nicholas, D. Bunandar, M. Pant, R. Steinbrecher Greg, J. Mower, M. Prabhu, T. Baehr-Jones, M. Hochberg, and D. Englund, "Large-scale quantum photonic circuits in silicon," in *Nanophotonics*(2016), p. 456.

152. A. Singh, Q. Li, S. Liu, Y. Yu, X. Lu, C. Schneider, S. Höfling, J. Lawall, V. Verma, R. Mirin, S. W. Nam, J. Liu, and K. Srinivasan, "Quantum frequency conversion of a quantum dot single-photon source on a nanophotonic chip," Optica **6**, 563-569 (2019)

153. J. Borregaard, A. S. Sørensen, and P. Lodahl, "Quantum Networks with Deterministic Spin–Photon Interfaces," Adv. Quantum Technol. **2**, 1800091 (2019)

154. L. M. Duan, and H. J. Kimble, "Scalable Photonic Quantum Computation through Cavity-Assisted Interactions," Phys. Rev. Lett. **92**, 127902 (2004)

155. N. Bar-Gill, L. M. Pham, A. Jarmola, D. Budker, and R. L. Walsworth, "Solid-state electronic spin coherence time approaching one second," Nat. Commun. **4**, 1743 (2013)

156. A. Lohrmann, B. C. Johnson, J. C. McCallum, and S. Castelletto, "A review on single photon sources in silicon carbide," Rep. Prog. Phys. **80**, 034502 (2017)

157. A. Gottscholl, M. Kianinia, V. Soltamov, C. Bradac, C. Kasper, K. Krambrock, A. Sperlich, M. Toth, I. Aharonovich, and V. Dyakonov, "Room Temperature Initialisation and Readout of Intrinsic Spin Defects in a Van der Waals Crystal," arXiv preprint arXiv:1906.03774 (2019)

158. H. Kim, R. Bose, T. C. Shen, G. S. Solomon, and E. Waks, "A quantum logic gate between a solid-state quantum bit and a photon," Nat. Photonics **7**, 373 (2013)

159. S. Sun, H. Kim, Z. Luo, G. S. Solomon, and E. Waks, "A single-photon switch and transistor enabled by a solid-state quantum memory," Science **361**, 57-60 (2018)

160. M. Heuck, K. Jacobs, and D. R. Englund, "Photon-Photon Interactions in Dynamically Coupled Cavities," arXiv preprint arXiv:1905.02134 (2019)

161. A. Faraon, I. Fushman, D. Englund, N. Stoltz, P. Petroff, and J. Vučković, "Coherent generation of non-classical light on a chip via photon-induced tunnelling and blockade," Nat. Phys. **4**, 859 (2008)

162. A. Javadi, I. Söllner, M. Arcari, S. L. Hansen, L. Midolo, S. Mahmoodian, G. Kiršanskè, T. Pregnolato, E. H. Lee, J. D. Song, S. Stobbe, and P. Lodahl, "Single-photon non-linear optics with a quantum dot in a waveguide," Nat. Commun. **6**, 8655 (2015)

163. D. Buterakos, E. Barnes, and S. E. Economou, "Deterministic Generation of All-Photonic Quantum Repeaters from Solid-State Emitters,"  Phys. Rev. X **7**, 041023 (2017)

164. B. Hacker, S. Welte, G. Rempe, and S. Ritter, "A photon–photon quantum gate based on a single atom in an optical resonator,"  Nature **536**, 193 (2016)

165. A. P. Foster, D. Hallett, I. V. Iorsh, S. J. Sheldon, M. R. Godsland, B. Royall, E. Clarke, I. A. Shelykh, A. M. Fox, M. S. Skolnick, I. E. Itskevich, and L. R. Wilson, "Tunable Photon Statistics Exploiting the Fano Effect in a Waveguide,"  Phys. Rev. Lett. **122**, 173603 (2019)

166. A. Osada, Y. Ota, R. Katsumi, M. Kakuda, S. Iwamoto, and Y. Arakawa, "Strongly Coupled Single-Quantum-Dot--Cavity System Integrated on a CMOS-Processed Silicon Photonic Chip,"  Phys. Rev. Appl. **11**, 024071 (2019)

167. G. Son, S. Han, J. Park, K. Kwon, and K. Yu, "High-efficiency broadband light coupling between optical fibers and photonic integrated circuits," in *Nanophotonics*(2018), p. 1845.

168. G. Reithmaier, S. Lichtmannecker, T. Reichert, P. Hasch, K. Müller, M. Bichler, R. Gross, and J. J. Finley, "On-chip time resolved detection of quantum dot emission using integrated superconducting single photon detectors,"  Sci. Rep. **3**, 1901 (2013)

169. F. Marsili, V. B. Verma, J. A. Stern, S. Harrington, A. E. Lita, T. Gerrits, I. Vayshenker, B. Baek, M. D. Shaw, R. P. Mirin, and S. W. Nam, "Detecting single infrared photons with 93% system efficiency,"  Nat. Photonics **7**, 210 (2013)

170. F. Najafi, J. Mower, N. C. Harris, F. Bellei, A. Dane, C. Lee, X. Hu, P. Kharel, F. Marsili, S. Assefa, K. K. Berggren, and D. Englund, "On-chip detection of non-classical light by scalable integration of single-photon detectors,"  Nat. Commun. **6**, 5873 (2015)

171. M. Widmann, M. Niethammer, T. Makino, T. Rendler, S. Lasse, T. Ohshima, J. U. Hassan, N. T. Son, S.-Y. Lee, and J. Wrachtrup, "Bright single photon sources in lateral silicon carbide light emitting diodes,"  Appl. Phys. Lett. **112**, 231103 (2018)

172. X. Lin, X. Dai, C. Pu, Y. Deng, Y. Niu, L. Tong, W. Fang, Y. Jin, and X. Peng, "Electrically-driven single-photon sources based on colloidal quantum dots with near-optimal antibunching at room temperature,"  Nat. Commun. **8**, 1132 (2017)

173. J. P. Lee, E. Murray, A. J. Bennett, D. J. P. Ellis, C. Dangel, I. Farrer, P. Spencer, D. A. Ritchie, and A. J. Shields, "Electrically driven and electrically tunable quantum light sources,"  Appl. Phys. Lett. **110**, 071102 (2017)

174. P. Munnelly, T. Heindel, A. Thoma, M. Kamp, S. Höfling, C. Schneider, and S. Reitzenstein, "Electrically Tunable Single-Photon Source Triggered by a Monolithically Integrated Quantum Dot Microlaser,"  ACS Photonics **4**, 790-794 (2017)

175. A. Dietrich, M. Doherty, I. Aharonovich, and A. Kubanek, "Persistence of Fourier Transform limited lines from a solid state quantum emitter in hexagonal Boron Nitride,"  arXiv preprint arXiv:1903.02931 (2019)

176. D. J. Brod, E. F. Galvão, A. Crespi, R. Osellame, N. Spagnolo, and F. Sciarrino, "Photonic implementation of boson sampling: a review,"  Adv. Photon. **1**, 1-14, 14 (2019)

177. P. Sibson, C. Erven, M. Godfrey, S. Miki, T. Yamashita, M. Fujiwara, M. Sasaki, H. Terai, M. G. Tanner, C. M. Natarajan, R. H. Hadfield, J. L. O'Brien, and M. G. Thompson, "Chip-based quantum key distribution,"  Nat. Commun. **8**, 13984 (2017)

178. J. Preskill, "Quantum Computing in the NISQ era and beyond,"  Quantum **2**, 79 (2018)